\documentclass[
aps,%
11pt,%
final,%
notitlepage,%
oneside,%
twocolumn,%
nobibnotes,%
nofootinbib,%
superscriptaddress,%
noshowpacs,%
centertags]%
{revtex4}

\begin{document}
%\selectlanguage{russian}
%\selectlanguage{english}

% сноски к author и affiliation:
% \email[E-mail:]{e-mail адрес}
% \homepage[]{}% web-страница
% \thanks{}% любой текст
% \noaffiliation % если не надо указывать место работы автора
% \altaffiliation{Второе место работы}
% \collaboration{ }

%\date{\today}

\title{RATAN-600 7.6-cm Deep Sky Strip Surveys at the Declination
of the SS433 Source During the 1980--1999 Period. Data Reduction
and the Catalog of Radio Sources in the Right-Ascension Interval
 $7^h \le$  R.A. $ < 17^h$}

\author{\firstname{N.~S.}~\surname{Soboleva}}
\affiliation{St. Petersburg Branch of the Special Astrophysical
Observatory, Russian Academy of Sciences, Pulkovo, St. Petersburg,
196140 Russia}

\author{\firstname{E.~K.}~\surname{Majorova}}
\affiliation{\saoname}

\author{\firstname{O.~P.}~\surname{Zhelenkova}}
\affiliation{\saoname}

\author{\firstname{A.~V.}~\surname{Temirova}}
\affiliation{St. Petersburg Branch of the Special Astrophysical
Observatory, Russian Academy of Sciences, Pulkovo, St. Petersburg,
196140 Russia}

\author{\firstname{N.~N.}~\surname{Bursov}}
\affiliation{\saoname}

%\received{April 16, 2009}%
%\revised{June 15, 2009}%

\begin{abstract}
We use two independent methods to reduce the data of the surveys
made with RATAN-600 radio telescope at 7.6\,cm in 1988--1999 at
the declination of the SS433 source. We also reprocess the data of
the ``Cold'' survey (1980--1981). The resulting RCR (RATAN COLD
REFINED) catalog contains the right ascensions and fluxes of
objects identified with those of the  NVSS catalog in the
right-ascension interval $7^h \le$  R.A. $ < 17^h$. We obtain the
spectra of the radio sources and determine their spectral indices
at 3.94 and 0.5\,GHz. The spectra are based on the data from all
known catalogs available from the  CATS, Vizier, and NED
databases, and the flux estimates inferred from the maps of the
VLSS and GB6 surveys. For 245 of the 550 objects of the  RCR
catalog the fluxes are known at two frequencies only: 3.94\,GHz
(RCR) and 1.4\,GHz (NVSS). These are mostly sources with fluxes
smaller than  30\,mJy. About  65\% of these sources have flat or
inverse spectra ($\alpha > -0.5$). We analyze the reliability of
the results obtained for the entire list of objects and construct
the histograms of the spectral indices and fluxes of the sources.
Our main conclusion is that all 10--15\,mJy objects found in the
considered right-ascension interval were already included in the
decimeter-wave catalogs.

\end{abstract}
%\pacs{95.80.+p, 95.85.Bh, 98.70.Dk}

\maketitle
%\textmakefnmark{0}{}

{\footnotesize {\bf DOI:}10.1134/S1990341310010050}

\section{INTRODUCTION}

In 1991 by Parijskij et al.~\cite{pa1:Soboleva_n} a catalog of
7.6-cm radio sources in the \mbox{$4^h \le$ R.A. $< 22^h$}
interval of right ascensions (RC-catalog) and 31-cm radio sources
in the \mbox{$4^h \le$ R.A. $< 13^h$} interval at the declination
of \mbox{$Dec_{2000}=5\degr \pm 20'$} was published. This catalog
was based on the results of observations made in \mbox{1980--1981}
with the RATAN-600 radio telescope in the meridian and azimuth of
$30\degr$ \cite{h:Soboleva_n,ber:Soboleva_n,pa2:Soboleva_n}. The
coordinate calibration was based on the then most accurate UTRAO
(365 MHz) catalog\footnote{The data for the sky strip of interest
was kindly provided to us by Prof.~J.~N.~Douglas prior to
publication.}.

After the publication of the NVSS and FIRST catalogs (1.4 GHz, VLA) \cite{co1:Soboleva_n,fr:Soboleva_n} the objects of the RC
catalog were compared with the objects of the former two catalogs and  20--25\% of RC objects
proved to be impossible to cross identify with NVSS objects \cite{zh:Soboleva_n}.

Several additional observation sets were carried out from 1987
through 1999 at the Northern sector of RATAN-600 in order to
refine the RC catalog and, in particular, the fluxes and
coordinates of its sources. The observations, like earlier, were
made at the declination of  SS433. The declination varied from
cycle to cycle because of precession. However, these variations
proved to be too small to allow the declinations of RCR objects to
be found with sufficient accuracy. The results of the reduction of
these observations in the right-ascension interval \mbox{$2^h \le
{\rm R.A.} < 7^h$} and \mbox{$17^h \le {\rm R.A.} < 22^h$} can be
found in our earlier paper~\cite{so1:Soboleva_n}.

In this paper we report the results of the reduction of the 7.6-cm
observations made in  1987--1999 in the right-ascension band $7^h
\le {\rm R.A.}< 17^h$. In addition, we also report  the results of
our rereduction of the records obtained in the ``Cold'' experiment
in 1980.

We uses NVSS objects to calibrate the right ascensions. We also
used the declinations of sources from NVSS catalog.

Almost the entire observed region has been studied with a VLA with a resolution of  5.4$''$
(the FIRST catalog, $ 8^{h}11^{m} \le {\rm R.A.} < 16^{h}26^{m}$), and the results of these observations allowed us
to refine the structure of the radio sources.

We report the list of objects (the RCR catalog) found within the right-ascension band mentioned above
and identified with  NVSS objects~\cite{co1:Soboleva_n}. We separately discuss the reliability of the identification of
our objects with the objects of the NVSS catalog and the statistical conclusions based on the rereduced RC catalog.
We pay special attention to objects with peculiar spectra and to objects with fluxes known only at two
frequencies: \mbox{1.4\,GHz} (NVSS) and 3.94 GHz (RCR).

\section{REDUCTION OF OBSERVATIONAL DATA}

The shape of the \mbox{RATAN-600} beam pattern differs
substantially from that of a paraboloid
antenna~\mbox{\cite{e1:Soboleva_n,e2:Soboleva_n,e3:Soboleva_n,m1:Soboleva_n,m2:Soboleva_n}}.
The width of the horizontal section of the beam extends with the
distance in declination from the beams center. Therefore we used
two different methods for data reduction, which differed mostly by
the substraction of the background on averaged scans.

We computed the background with a  80 and 20-s ``smoothing
window'' in the first and second methods, respectively. We chose
the ``smoothing window'' based on the computed $HPBW(\Delta Dec)$
dependences, which were tested \mbox{experimentally
\cite{m2:Soboleva_n,m3:Soboleva_n}} ($HPBW$ is the beam halfwidth
in the horizontal section and $\Delta Dec$ is declination
difference between the section considered and the central
horizontal section).

It is safe to conclude that the results obtained with the
background computed  and then substracted with a 80-s and 20-s
``window'' should filter out the radio sources that cross the beam
pattern at declinations outside the $Dec_{0} \pm 50'$ and $Dec_{0}
\pm 12'$ bands, where $Dec_{0}$ is the declination of the central
section of the observation set.

In practice, after the substraction of the background computed
with a 20-s ``window'' the sources at $\Delta Dec > \pm 8'$ are
not excluded completely, but only decreased their antenna
temperature. No reduction of the antenna temperatures is observed
for sources in the  $\Delta Dec \le \pm 8'$ band.

The effect of the atmosphere may show up if the background is
computed with a 80-s ``window'', whereas such effects are minimal
if the background is computed with a 20-s wide ``window'', because
in this case the low-frequency noise---and atmospheric
fluctuations in particular---is filtered out.

Thus the two methods result in different systematic errors in the
inferred fluxes of the radio sources. Whereas the first method may
overestimate the fluxes caused by the influence of the atmosphere,
the second method tends to underestimate the fluxes of distant
objects.

After substraction the background level we extracted the sources
from averaged scans by applying Gaussian analysis. We performed
R.A. calibration using strong sources and data of the NVSS
catalog. For each identified source we computed its  R.A.
coordinates, antenna temperature (${\rm T_{a}}$), and the
halfwidth of the Gaussian ($HPBW_{i}$). In addition in the first
method we drew the ``zero'' line within the given object before
fitting the Gaussian.

\subsection{First Method}

One-dimensional averaged scans are superpositions of all sources
that have crossed different horizontal sections of the beam
pattern, and therefore to discriminate and identify these sources
more reliably we simulated the survey using NVSS
images~\mbox{\cite{co1:Soboleva_n,na:Soboleva_n}} in the same way
as Majorova~\cite{m4:Soboleva_n} did it for the \mbox{RZF survey
\cite{bu:Soboleva_n}.}

\begin{figure*}[tbp]
\onelinecaptionsfalse
\centerline{
%\vbox{
\hbox{
\includegraphics[angle=0,width=0.5\textwidth,clip]{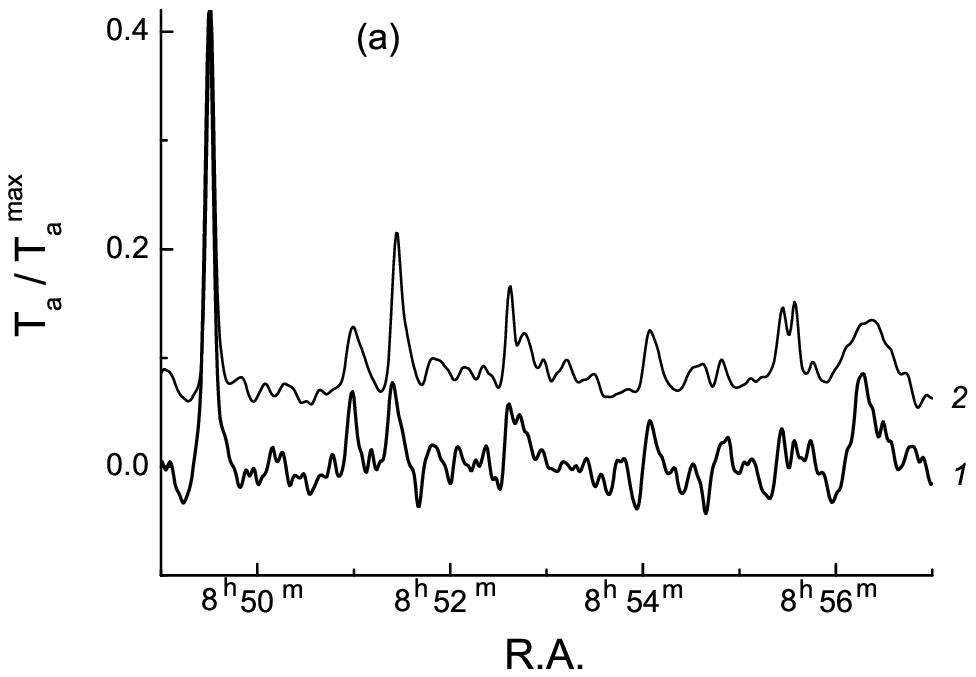}
%}
%\hbox{
\includegraphics[angle=0,width=0.5\textwidth,clip]{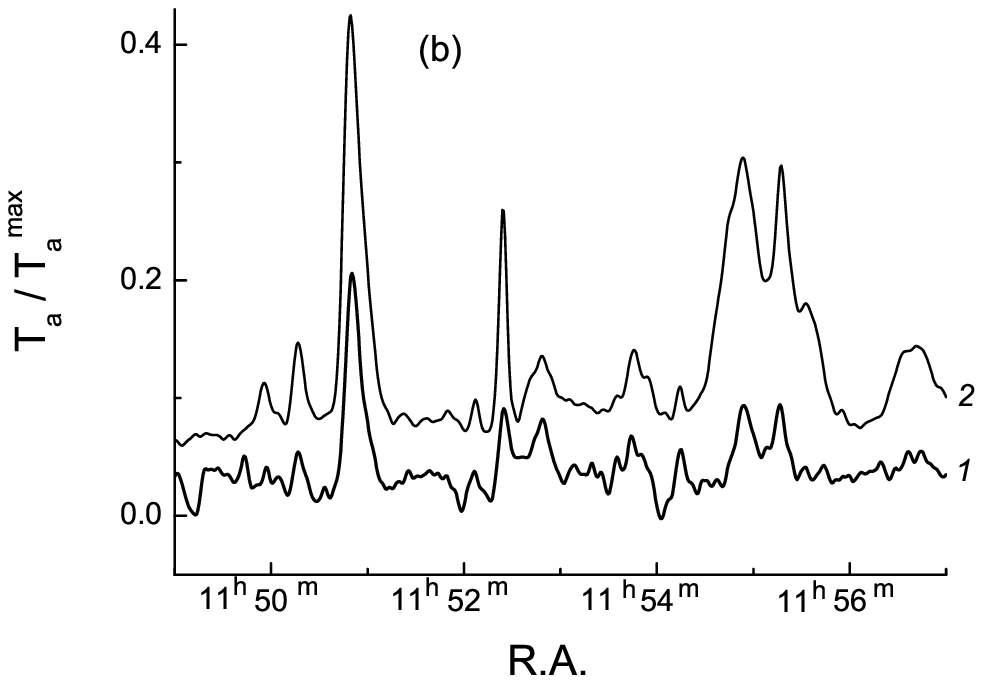}
}
}
%}
\setcaptionmargin{5mm}
\captionstyle{normal}
\caption{
Normalized real averaged records of the RCR survey at 7.6~cm (the curves {\it 1}) and normalized
simulated scans of the same sky areas based on NVSS images (the curves {\it 2}). The scans in Figs.~1a and 1b
are normalized to the level of the signal from the sources 082056+045417 and 114520+045526, respectively.
The y-axis is in relative units and the x-axis gives the right ascension.
}
\label{fig0:Soboleva_n}
\end{figure*}

We then compared the normalized simulated scans with real
7.6\,cm-records after the substraction of the 80-s background.
Figure\,\ref{fig0:Soboleva_n} demonstrates examples of normalized
simulated and real averaged scans. The scans in
Figs.~\ref{fig0:Soboleva_n}a and~\ref{fig0:Soboleva_n}b are
normalized to the level of the signal from the sources
082056+045417 and 114520+045526, respectively. The y-axis is in
relative units and the x-axis gives the right ascension. Simulated
scans proved to be most efficient in complex cases involving the
superposition of two or more sources.

%Fig 1

When reducing the data for each observation set we used the
$\Delta Dec$ values of the NVSS sources located within the
\mbox{$Dec_{0} \pm 1\degr$} band and the \linebreak $HPBW(\Delta
Dec)$ dependences. We also computed the pattern factor
\mbox{$k_{DN}(\Delta Dec)$} and the product \mbox{${\rm
F_{1.4}}\times k_{DN}(\Delta Dec)$}, where ${\rm F_{1.4}}$ is the
source flux at \mbox{1.4\,GHz.} We computed $HPBW(\Delta Dec)$ and
$k_{DN}(\Delta Dec)$ using the algorithms published by
Majorova~\cite{m1:Soboleva_n}. The pattern factor meant for the
reduction of the response from the source as a function of its
distance from the central section  of the survey (or the central
section of the beam pattern). This factor is used to find the
fluxes of the sources. The product \mbox{${\rm F_{1.4}}\times
k_{DN}(\Delta Dec)$} provides information about the probability of
the source appearance on the real scan.  This product also helps
in extraction and identification of the sources with close right
\mbox{ascensions.}

For further analysis we selected from the list of NVSS sources
located within the $Dec_{0} \pm 1\degr$ \linebreak band those that
satisfied the condition \linebreak \mbox{${\rm F_{1.4}}\times
k_{DN}(\Delta Dec)
> 3$}.
Our subsequent investigation showed that no sources with
\linebreak \mbox{${\rm F_{1.4}}\times k_{DN}(\Delta Dec) < 3$} had
been found in the records.

A comparison of the $HPBW$ values inferred via Gaussian analysis
with the computed values allowed us to control the reliability of
the extraction of the sources and, in a number of cases, helped us
to understand whether we found the right source in the record.

Thus, in the first method of data reduction we use three criteria
to control the extraction of the sources: closeness of the R.A. of
the source in the record to the  R.A. of the source according to
the \mbox{NVSS catalog,} closeness of the halfwidth of the
Gaussian to its computed value, and sufficiently large value of
the product ${\rm F_{1.4}}\times k_{DN}(\Delta Dec)$. The
sensitivity varied from cycle to cycle and therefore the minimum
(threshold) product also differed from cycle to cycle.

In the first method we computed the source fluxes ${\rm F_{1}}$ by the following formula:

    ${\rm F_{1}} = A\times {\rm T_{a}}/k_{DN}$,

$A=2k/S_{eff}$, where $k$ is the Boltzmann constant and $S_{eff}$ is the effective area of the radio telescope.

We found the coefficient $A$ individually for each observation set
using the technique described by Majorova and Bursov~\cite{m3:Soboleva_n}. To
this end, we selected sufficiently strong sources with steep
spectra and known spectral indices, computed their fluxes at
7.6~cm, and determined from the real scans the antenna
temperatures of theses sources for  each observation set. We then
constructed the dependences ${\rm F/T_{a}}(\Delta Dec)$, that we
fitted by the appropriate approximating curves using the least
squares method. The coefficient $A$ is equal to the value of the
approximating curve at $\Delta Dec=0$ for the observation set
considered.

We computed the pattern factor $k_{DN}$ in accordance with the
$\Delta Dec$ of the source. In our computations we took into
account the offset of the primary horn along the focal line of the
secondary mirror in the observation set considered. We then
averaged the inferred fluxes over all observation sets. We list
the fluxes ${\rm F_{1}}$ and right ascensions R.A.$_{1}$ with
their errors in the Table.

Note that part of the sources could be found only in one of the
observation sets, in particular, in the 1980 set, that was
characterized by the highest sensitivity. In some of the cases
sources could be identified only in the records made in 1994, when
the antenna was mounted~$4'$ lower ($Н=51\degr 22'$) or higher
\mbox{($Н=51\degr 09'$)} than the declination of SS433. The errors
of the fluxes determined for sources identified in only one
observation set amounts to ($20\pm 5)\%$.

The yearly-averaged noise root-mean-squared errors in  $\overline{3\sigma}$ records are (in mK):
$ 2.2 \pm0.5$~mK for the 1980 set; \mbox{$3.4\pm0.4 $\,mK } for the 1988 set;
$  4.9\pm1.7 $ ~mK for the 1993 set; \mbox{$  3.2\pm0.4 $ mK} for the 1994 set; $  4.5\pm1.7 $
~mK for the 1994 set \mbox{($Н=51\degr 09'$),} and $  3.7\pm0.7 $ ~mK for the 1994 set
($Н=51\degr 22'$).

\begin{figure}[h]
\onelinecaptionsfalse
\centerline{
\hbox{
\centerline{
\includegraphics[width=7cm, bb=17 10 285 227, clip]{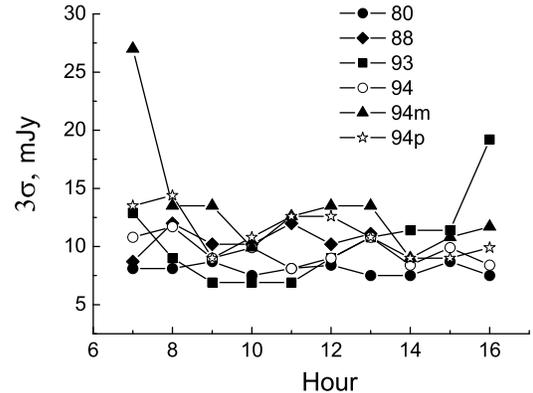}
} } } \setcaptionmargin{0mm} \captionstyle{normal}
\caption{Root-mean-squared noise error ($3\sigma$) of the averaged
scans as a function of the hour of observation computed for
observation sets of different years. Here 94m and 94p correspond
to the scans obtained in 1994 with the antenna mounted $4'$ below
and above the declination of SS433, respectively. The scans with
the 80-s background subtracted are used. } \label{fig01:Soboleva_n}
\end{figure}

Figure~\ref{fig01:Soboleva_n} demonstrates the variations of the
root-mean-squared noise error ($3\sigma$) (in mJy) for averaged
scans at different observation hours. Here 94m and 94p correspond
to the records obtained in 1994 with the antenna mounted $4'$
below and above the declination of SS433, respectively.

The mean $\overline{3\sigma}$ values (in mJy) averaged over all observations are equal to: $ 8.0 \pm0.5 $~mJy
for the  \mbox{1980} set;  $ 10.6\pm1.3 $ mJy for the 1988 set; \mbox{$ 10.4\pm3.7 $ mJy} for the \mbox{1993}
set; $  9.6\pm1.2 $~mJy for the \mbox{1994} set;  \mbox{$ 13.5\pm5.5 $~mJy} for the \mbox{1994}
($Н=51\degr 09'$) set, and  $ 11.1\pm2.0 $~mJy for the \mbox{1994}  \mbox{($Н=51\degr 22'$) set.} The
root-mean-squared errors were computed for the scans with the 80-s background subtracted.

%Fig.2

\subsection{Second Method}

After substraction the 20-s background using the second method we
applied Gaussian analysis to find the positions of the radio
sources on the scan (R.A.), their antenna temperatures, and
halfwidths. We then used the antenna temperatures to compute the
fluxes  ${\rm F_{2}}$ of the radio sources by the following
formula:

        ${\rm F_{2}} =k_{eff}\times k_{i}\times {\rm T_{a}}/k_{DN}$,

\noindent where $k_{eff}$ is the factor that takes into account
the effective area of the antenna and $k_{i}$ is the correcting
factor to correct for the differences between the calibrations and
small differences between the effective antenna areas during
different observation sets. We determined  $k_{i}$ from the
sources with well-known spectra found within the survey band. The
$k_{i}$ factors lied in the  \mbox{1.1--1.5} interval depending on
the year of observation. The $k_{eff}$ factor was equal to 3.5. We
computed the pattern factor for each radio source with the
transversal offset of the primary feed along the focal line of the
secondary mirror and the vertical distance of the radio object
from the central section of the beam pattern taken into account in
accordance with the algorithms described by
Majorova~\cite{m1:Soboleva_n}.

Our significance criterion is based on the presence of the object
on the scans of more than two observation sets. Prior to
identification of objects we smoothed the scans by applying the
computed beam pattern in its central section. A detailed
description of the technique can be found
in~\cite{so2:Soboleva_n}.

After finding the fluxes and right ascensions in each observation
set we computed the flux ${\rm F_{2}}$ and R.A.$_{2}$ averaged
over all observation runs and the corresponding errors (see
Table).

In this method of data reduction the noise root-mean-squared error in records is somewhat lower than in
the first method. This is due, first, to smoothing of records by the beam and, second, to  ``trimming'' of
noise after the subtraction of the 20-s background.

\section{CATALOG OF RADIO SOURCES AT 7.6\,CM}

The inferred 7.6~cm fluxes and right ascensions of the sources are given in the catalog whose fragment we
list in the Table.

Column~1 gives the J2000.0 coordinates of the sources according to
the NVSS catalog; columns~2 and 3, the right-ascension differences
of the objects between the NVSS coordinates and our estimates
${\rm \Delta RA_{1}}$ (the first method proposed by Majorova) and
$\Delta {\rm RA}_{2}$ (the second method proposed by Soboleva and
Temirova); columns~4 and 5, the fluxes with the corresponding
errors: ${\rm F_{1}}$, the fluxes found by E.~K.~Majorova using
the first method, and ${\rm F_{2}}$, the fluxes found by
N.~S.~Soboleva and A.~V.~Temirova using the second method.
Columns~6 and 7 give the spectral indices $\alpha_{3.94}$ and
$\alpha_{0.5}$ ($F_{\nu} \sim \nu^{\alpha}$) of the objects found
at two frequencies 3.94 and 0.5~GHz (we chose the 0.5 GHz
frequency by analogy with~\cite{mi:Soboleva_n}). Columns~8--10 contain the
comments.

We used the source fluxes determined using the two methods to construct the spectra. To this end, we also used
all the known catalogs available via the  CATS \cite{v1:Soboleva_n,v2:Soboleva_n}, Vizier \cite{viz:Soboleva_n}, and NED \cite{ned:Soboleva_n} databases,
which intersect with the band of our sky survey. We approximated the spectra by the curves used to determine
the spectral indices of the objects at 3.94 ($\alpha_{3.94}$) and 0.5~GHz ($\alpha_{0.5}$).

When constructing the spectra we also used, in addition to the
available data from known catalogs, the flux estimates obtained by
O.~P.~Zhelenkova based on the maps of the VLSS (74~MHz)
\cite{cohen:Soboleva_n} and GB6 (4850~MHz) \cite{gregory:Soboleva_n} surveys. These
estimates were obtained for the objects of our list missing from
the VLSS and GB6 catalogs, because they had fluxes below the $
5\sigma$ level \mbox{(here $\sigma$ } is the r.m.s. error of noise
for the given map). We extracted objects with fluxes $3\sigma \le
$ F < $ 5\sigma$, that are marked by letters V and G in column~8
of the Table.

Furthermore, we also marked the sources with spectral
peculiarities (hill, GPS (Gigahertz Peak Spectrum), HFP (High
Frequency Spectrum), and var (variable)), double sources (d), and
unresolved sources (b), i.e., blends. The double asterisk (**)
indicates large scatter of the spectral data for the source; hill,
HFP, and GPS indicate the objects that exhibit a maximum in the
spectrum---in particular, ``hill'' points to a source with a small
maximum in the spectrum at a frequency close to  3.94~GHz. The
asterisk~(*) in column~9 indicates objects with antenna
temperatures in the records lie in the \mbox{3$\sigma \le T_{a} <
5 \sigma$} interval. The antenna temperatures of all other sources
exceed $5 \sigma$.

%\clearpage
%\newpage
%\onecolumngrid

%\begin{center}
\setcaptionmargin{5mm}
\onelinecaptionsfalse
\captionstyle{nonumber}
\begin{longtable*}{c|l|l|l|l|l|l|l|l|l}
\caption{{\bf Table.} A fragment of the  RCR catalog at 7.6~cm in
the right-ascension interval  \mbox{$7^h \le$ R.A. $ < 17^h$}
(1980--1999)} \\
\hline
~${\rm R.A.}_{2000}$~~~$Dec_{2000}$   &~$\Delta {\rm RA}_{1}\pm\sigma$&~$\Delta {\rm RA}_{2}\pm\sigma$&~${\rm F_{1}}\pm\sigma$&~${\rm F_{2}}\pm\sigma$&~$\alpha_{3.94}$&~~$\alpha_{0.5}$&pr1&pr2&pr3\\
 ~~NVSS      ~  &~~s.ss~~~~s.ss      ~&~~s.ss~~~~s.ss      ~&~~      mJy      &~~     mJy       &                &                &    &   &   \\
\hline
  (1)             &~~~~~~~~~~(2)                &~~~~~~~~~~(3)                &~~~~~(4)         &~~~~~(5)         &~~(6)           &~~(7)          & ~(8)      &~(9)  &~(10)\\
\hline
\endfirsthead

\hline
  (1)             &~~~~~~~~~~(2)                &~~~~~~~~~~(3)                &~~~~~(4)         &~~~~~(5)         &~~(6)           &~~(7)          & ~(8)      &~(9)  &~(10)\\
\hline
\endhead

\hline
\endfoot

\hline
\endlastfoot

~ 070209.13+044011.3  &~  1.03 $\pm$0.51 &~ --1.97  $\pm$0.84 &~247~$\pm$25  &~174~$\pm$10  &~--0.98~&~--0.72~&~      &~  &    \\
~ 070210.69+044837.1  &~  2.59 $\pm$0.50 &~                  &~19.4$\pm$1.4 &              &~--0.68~&~     ~&~      &~  &\#  \\
~ 070309.94+045510.7  &~  1.00 $\pm$0.42 &~  0.70  $\pm$0.88 &~13.1$\pm$3.0 &~14.2$\pm$2.0 &~--0.61~&~     ~&~      &~* &\#  \\
~ 070424.60+050418.7  &~ --0.31 $\pm$0.36 &~  0.66  $\pm$0.27 &~26.6$\pm$4.0 &~19.3$\pm$3.3 &~--0.30~&~     ~&~      &~* &    \\
~ 070451.88+050358.9  &~  0.00 $\pm$0.73 &~  0.06  $\pm$0.94 &~18.4$\pm$3.5 &~24.3$\pm$2.4 &~ 0.01~&~     ~&~d   G &~* &\#  \\
~ 070612.43+045546.8  &~  0.33 $\pm$0.54 &~  0.14  $\pm$0.41 &~~8.2$\pm$1.2 &~~9.3$\pm$2.1 &~--0.34~&~     ~&~      &~* &\#  \\
~ 070745.79+045525.9  &~                 &~                  &              &              &       &~     ~&~b     &~* &\#  \\
~ 070747.30+045414.0  &~  0.92 $\pm$0.43 &~  0.51  $\pm$0.10 &~21.7$\pm$2.1 &~22.9$\pm$2.7 &~--0.19~&~     ~&~b   G &~  &\#  \\
~ 071130.04+045140.0  &~  0.04 $\pm$0.30 &~  0.42  $\pm$0.32 &~33.4$\pm$3.0 &~39.0$\pm$3.9 &~--0.79~&~     ~&~b   G &~  &\#  \\
~ 071130.37+050037.0  &~  0.37 $\pm$0.30 &~  0.75  $\pm$0.32 &              &              &       &~     ~&~b     &~  &\#  \\
~ 071141.16+045416.2  &~ --0.29 $\pm$0.27 &~  0.08  $\pm$0.80 &~~8.3$\pm$1.4 &~10.0$\pm$1.0 &~--0.20~&~     ~&~      &~* &\#  \\
~ 071350.65+050210.2  &~  0.53 $\pm$0.25 &~  0.01  $\pm$0.13 &~29.8$\pm$4.8 &~23.2$\pm$4.2 &~--0.69~&~--0.69~&~ V    &~  &    \\
~ 071414.01+045526.9  &~  0.54 $\pm$0.10 &~                  &~11.0$\pm$6.0 &~             &~--0.57~&~     ~&~ G    &~* &\#  \\
~ 071616.25+050016.3  &~ --0.65 $\pm$0.51 &~  2.19  $\pm$0.75 &~10.0$\pm$1.5 &~~9.7$\pm$1.8 &~--0.19~&~     ~&~ G    &~  &\#  \\
~ 071833.98+045632.9  &~ --0.06 $\pm$0.48 &~  0.13  $\pm$0.35 &~35.7$\pm$5.0 &~20.5$\pm$4.5 &~--0.75~&~--0.75~&~b     &~  &    \\
~ 071834.62+045248.0  &~  0.58 $\pm$0.50 &~  0.77  $\pm$0.35 &              &              &       &~     ~&~b   G &   &    \\
~ 071900.99+044705.2  &~ --0.06 $\pm$1.40 &~  1.56  $\pm$0.71 &~51.4$\pm$6.7 &~38.4$\pm$5.3 &~--1.06~&~--1.06~&~      &~  &    \\
~ 071910.80+045746.6  &~  0.67 $\pm$0.32 &~  1.17  $\pm$0.45 &~~9.5$\pm$1.5 &~14.6$\pm$4.2 &~--0.52~&~     ~&~      &~* &\#  \\
~ 072043.38+045026.1  &~  2.06 $\pm$1.30 &~  0.26  $\pm$0.63 &~28.2$\pm$4.2 &~21.3$\pm$4.7 &~--0.27~&~     ~&~ G    &~  &\#  \\
~ 072219.15+045455.4  &~ --0.11 $\pm$0.31 &~ --0.16  $\pm$0.18 &~23.9$\pm$2.2 &~22.2$\pm$2.1 &~--0.29~&~     ~&~ G    &~  &\#  \\
~ 072318.92+045535.0  &~ --0.13 $\pm$0.37 &~ --0.03  $\pm$0.59 &~11.8$\pm$1.3 &~16.3$\pm$1.0 &~--0.91~&~--0.91~&~ G, V &~* &\#  \\
~ 072415.87+044525.1  &~  1.20 $\pm$0.83 &~  2.02  $\pm$0.61 &~78.6$\pm$2.8 &~60.5$\pm$3.1 &~--0.71~&~--0.87~&~      &~  &    \\
~ 072500.61+045008.9  &~ --0.16 $\pm$0.25 &~  0.34  $\pm$0.78 &~20.7$\pm$3.0 &~16.0$\pm$1.3 &~--0.37~&~     ~&~      &~* &\#  \\

\end{longtable*}
%\end{center}

The \# sign in column~10 indicates the objects with flux data
available only from two catalogs---NVSS and RCR (RC). Part of
these objects have flux estimates based on the VLSS and GB6 maps.

The 4850\,MHz flux estimates based on GB6 maps made it possible to
corroborate the reliability of the extraction of faint objects
from our scans. The 74-MHz flux estimates (VLSS maps) confirmed
the linear form of the spectra of radio sources in a number of
cases. However, we did not use these estimates to find the
spectral index at 500\,MHz for the sources with no flux data at
frequencies below 1.4\,GHz. For this reason, we do not give the
corresponding spectral indices $\alpha_{0.5}$ in the Table even
for objects with linear spectra.

We extracted a total of 550 objects in the right-ascension band
\mbox{$7^h \le {\rm R.A.} < 17^h$} including 18 blends and 15
double sources. All these objects are identified with NVSS
sources. For 245 sources the fluxes are known only at two
frequencies (3.94~GHz---RATAN-600 and 1.4~GHz---VLA ). Some of
them have been corroborated by 4.85-GHz flux estimates based on
GB6 maps or 7.7 and 11.111~GHz RATAN-600 data obtained during the
same observation sets.

The catalog has a gap in the region \linebreak
\mbox{$16^{h}50^{m}55^{s} < {\rm R.A.} <16^{h}51^{m}30^{s}$,}
since objects located in this area are blended with the bright
extended radio galaxy Hercules A. In addition, the RCR catalog has
about one-minute wide gaps at the beginning of each hour due to
the calibration of the radiometer.

\section{RELIABILITY OF THE RCR OBJECTS IDENTIFICATION}

To assess the reliability of the identification of RCR objects
with the objects of the NVSS catalog, we used the data for the
entire list of sources in the right-ascension range $7^h \le {\rm
R.A.} < 17^h$.

We computed the NVSS minus RCR right-ascension differences ${\rm
\Delta RA}_{1}$, ${\rm \Delta RA}_{2}$ for the sources and
constructed the corresponding histograms for the first and second
methods of data reduction and for the averaged  ${\rm \Delta
RA}_{mean}$. We present these histograms in Fig.~3.

Figures~3a--3c show the histograms of the ${\rm \Delta
RA}_{1}={\rm R.A.}_{NVSS}-{\rm R.A.}_{1}$, ${\rm \Delta
RA}_{2}={\rm R.A.}_{NVSS}-{\rm R.A.}_{2}$, and ${\rm \Delta
RA}_{mean} = {\rm R.A.}_{NVSS} - {\rm R.A.}_{mean}$
right-ascension differences, where, ${\rm R.A.}_{mean}= ({\rm
R.A.}_{1}+{\rm R.A.}_{2})/2$.

%Fig.3
\begin{figure}[tbp]
\vspace{1cm} \onelinecaptionsfalse \centerline{ \vbox{ \hbox{
\centerline{
%\bigskip
\includegraphics[width=8cm]{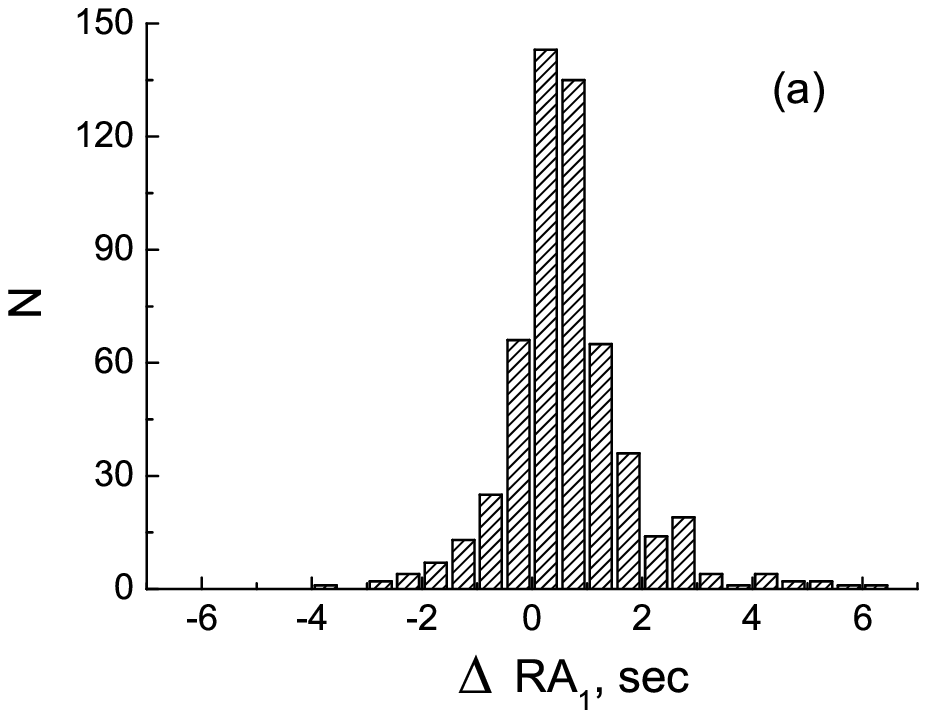}
}}
\hbox{
\centerline{
\includegraphics[width=8cm]{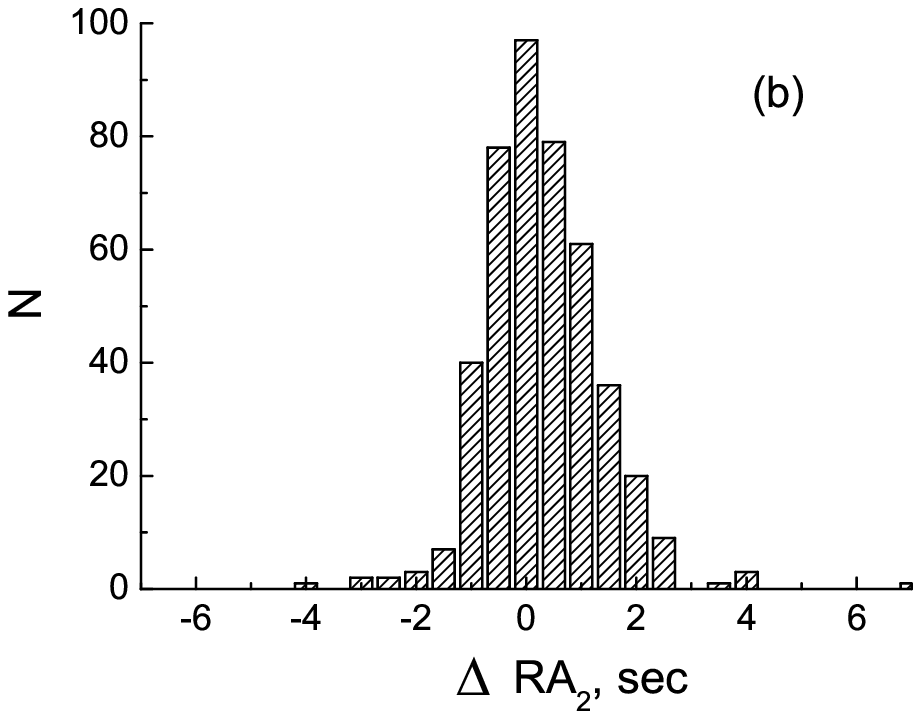}
}}
\hbox{
\centerline{
\includegraphics[width=8cm]{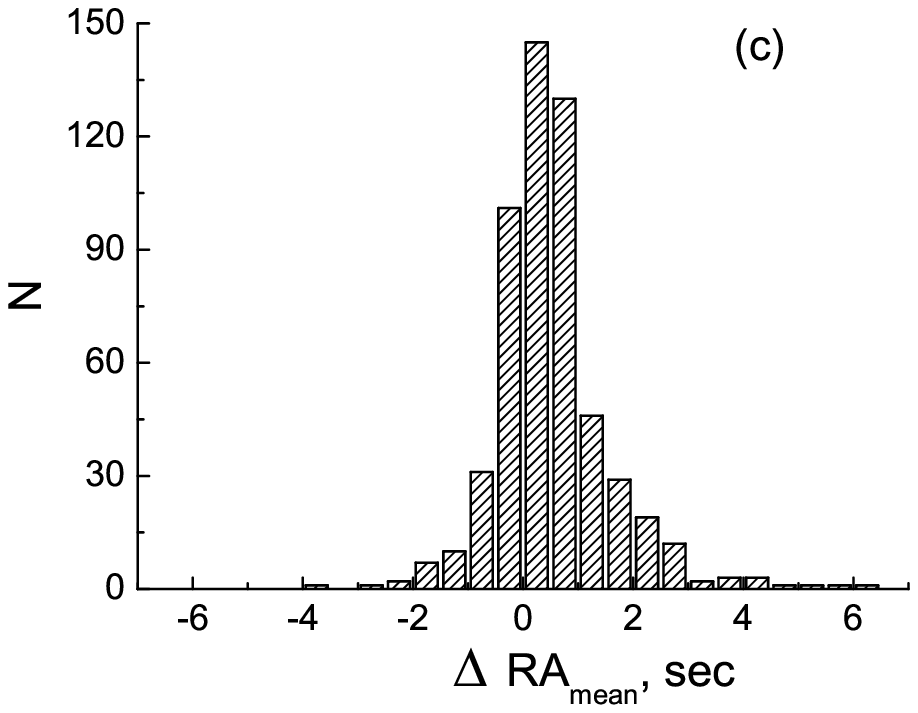}
} } } } \setcaptionmargin{0mm} \captionstyle{normal} \caption{NVSS
minus RCR right-ascension differences for RCR sources: \mbox{${\rm
\Delta RA = R.A.}_{NVSS}-{\rm R.A.}_{1}$ (\ref{fig1:Soboleva_n}a);}
\mbox{${\rm \Delta RA = R.A.}_{NVSS}-{\rm R.A.}_{2}$
(\ref{fig1:Soboleva_n}b);} \mbox{${\rm \Delta RA = R.A.}_{NVSS}-{\rm
R.A.}_{mean}$ (\ref{fig1:Soboleva_n}c),~where} \mbox{${\rm R.A.}_{mean}=({\rm
R.A.}_1+{\rm R.A.}_2)$/2.} } \label{fig1:Soboleva_n}
\end{figure}

%Fig.4
\begin{figure}[tbp]
\vspace{1cm} \onelinecaptionsfalse
%\centerline{
\vbox{
%\hbox{
%\centerline{
\includegraphics[width=8cm]{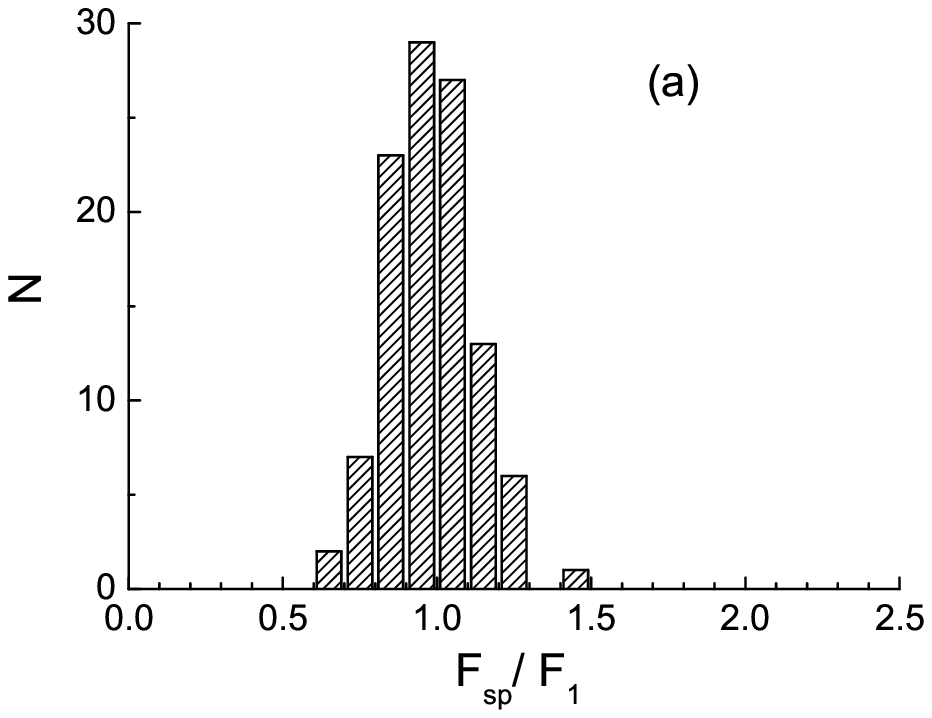}
%}}
%\hbox{
%\centerline{
\includegraphics[width=8cm]{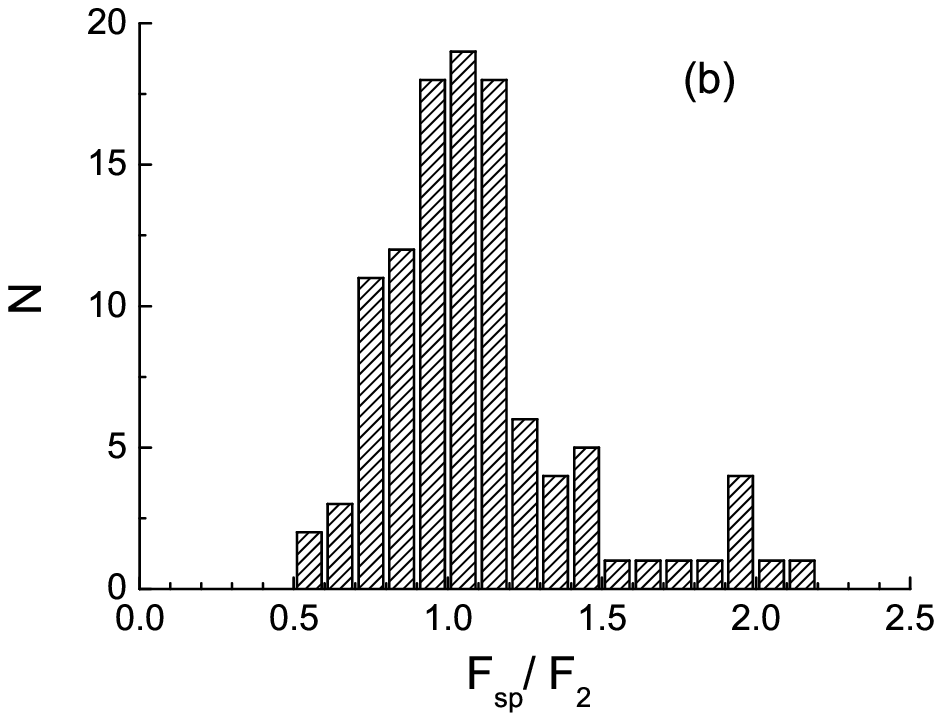}
%}}
%\hbox{
%\centerline{
\includegraphics[width=8cm]{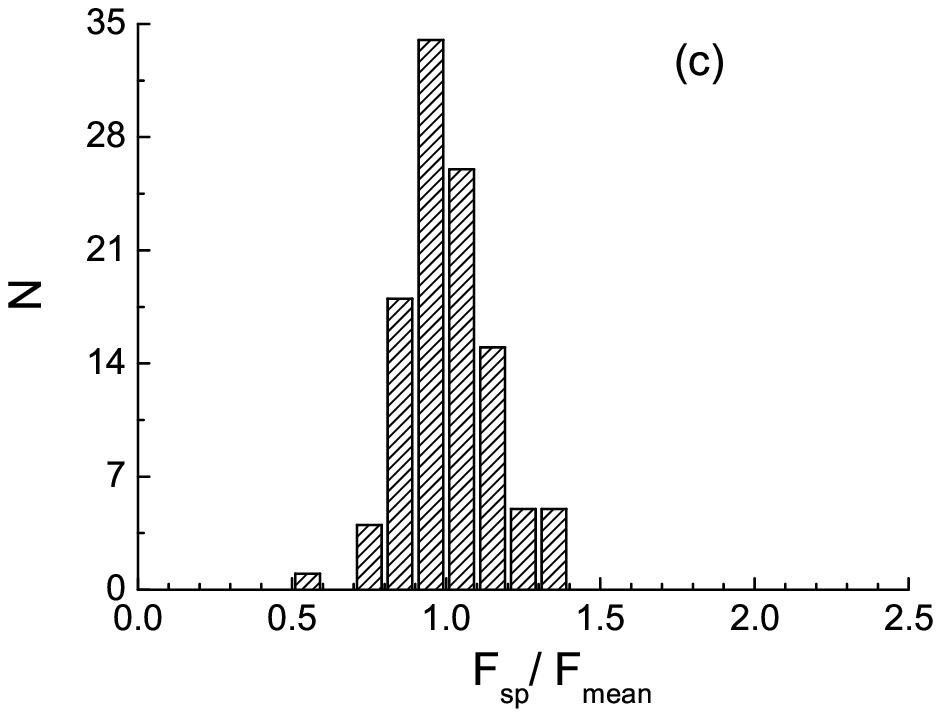}
%}
%}
}%}
\setcaptionmargin{0mm} \captionstyle{normal} \caption{Histograms
of the ratios of the expected 3.94~GHz fluxes $F_{sp}$ inferred
from the spectra of well studied sources to the fluxes of the same
sources determined in this paper using different methods. In fig.
\ref{fig2:Soboleva_n}a we present ${\rm F_{sp}/F_{1}}$, in \ref{fig2:Soboleva_n}b: ${\rm
F_{sp}/F_{2}}$, and in \ref{fig2:Soboleva_n}c: ${\rm F_{sp}/F_{mean}}$, where
\mbox{${\rm F_{mean}}=({\rm F}_1+{\rm F}_2)$/2.} } \label{fig2:Soboleva_n}
\end{figure}

%Fig.5
\begin{figure*}[tbp]
\onelinecaptionstrue
\centerline{
\hbox{
\centerline{
\includegraphics[angle=0,width=0.45\textwidth,clip]{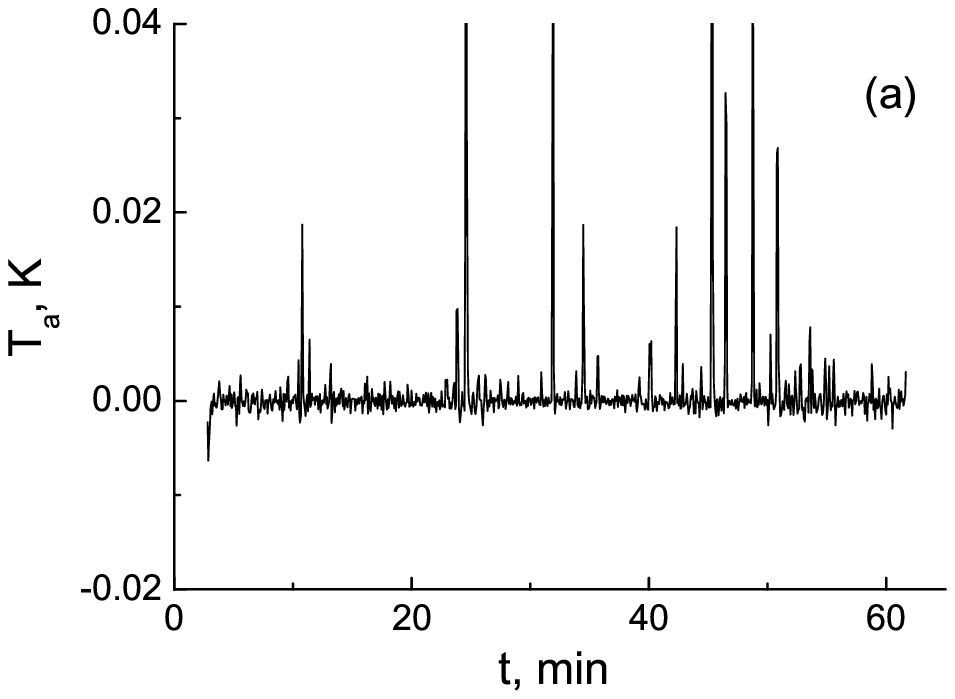}
\includegraphics[angle=0,width=0.45\textwidth,clip]{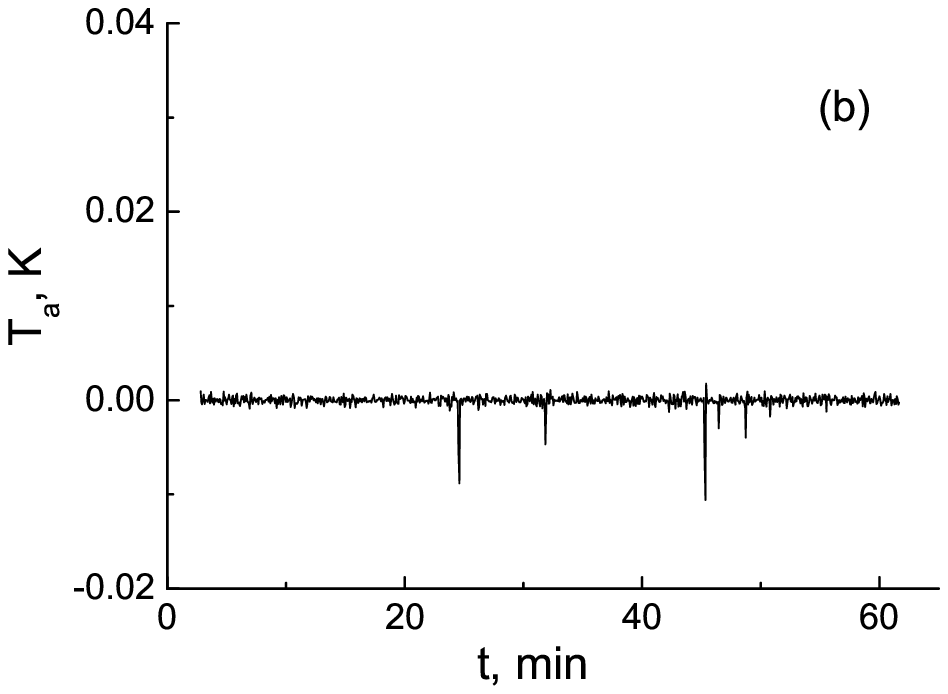}
%}
} } } \setcaptionmargin{0mm} \captionstyle{normal} \caption{ The
sum (\ref{fig16:Soboleva_n}a) and difference (\ref{fig16:Soboleva_n}b) of the averaged
scans obtained from two groups of 11-h records taken in 1994. }
\label{fig16:Soboleva_n}
\end{figure*}

%Fig.6
\begin{figure}[tbp]
\onelinecaptionsfalse
\centerline{
\vbox{
\hbox{
\centerline{
\includegraphics[width=8cm]{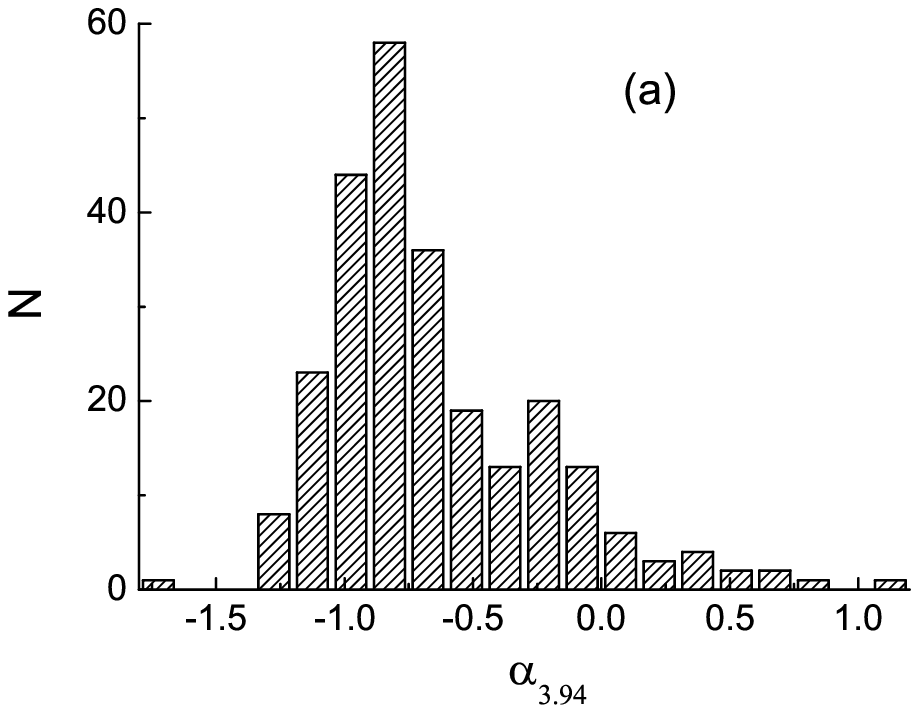}
}}
\hbox{
\centerline{
\includegraphics[width=8cm]{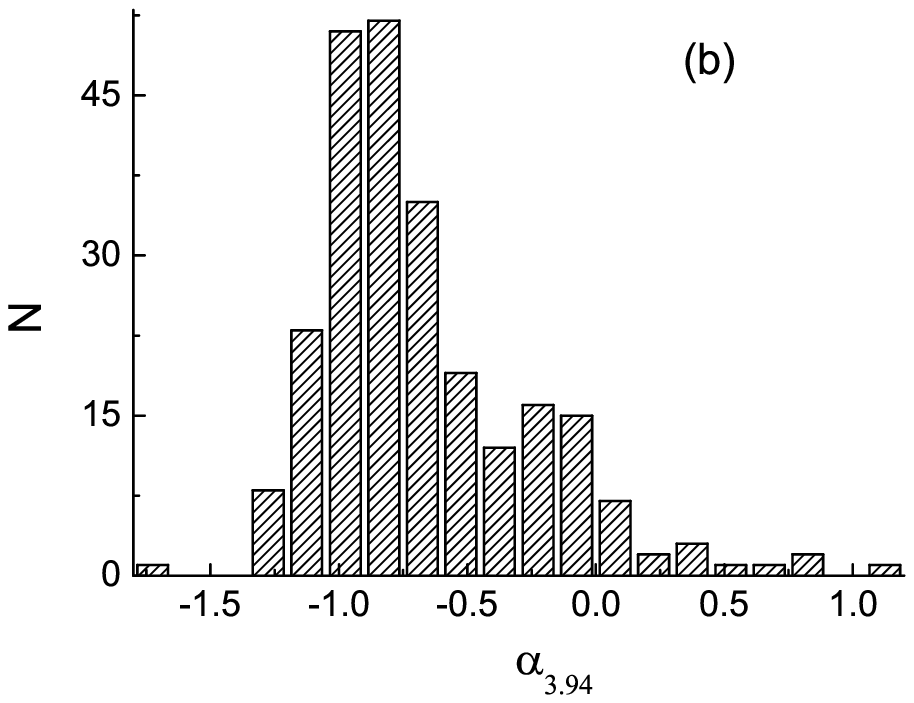}
}}
\hbox{
\centerline{
\includegraphics[width=8cm]{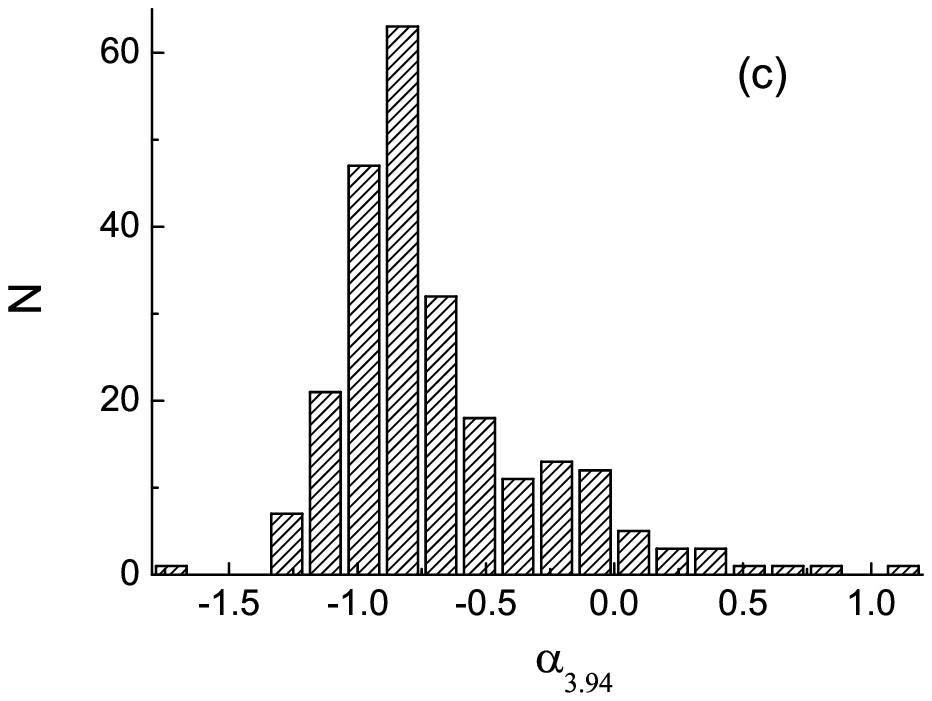}
}} } } \setcaptionmargin{0mm} \captionstyle{normal}
\caption{Histograms of spectral indices at 3.94~GHz for the
sources of the second group with well-studied spectra: computed
taking into account the estimates based on GB6 survey maps
(\ref{fig3:Soboleva_n}a), computed taking into account the estimates based on
VLSS maps (\ref{fig3:Soboleva_n}b), and computed taking into account the
estimates based on GB6 maps and selected estimates based on VLSS
maps (\ref{fig3:Soboleva_n}c). } \label{fig3:Soboleva_n}
\end{figure}

In addition, we estimated the mean $\overline{{\rm \Delta RA}}$
averaged over the entire sample of sources and the two methods of
data reduction. The mean $\overline{{\rm \Delta RA}}$ values
averaged over the entire sample \mbox{are equal to:} \\
$\overline{{\rm \Delta RA}_{1}} = (0.63\pm1.11$),s,
\\ \mbox{$\overline{{\rm \Delta RA}_{2}} = (0.29\pm1.08$)\,s,} and
\\ $\overline{{\rm \Delta RA}_{mean}} =  (0.52\pm1.05$)\,s.

For the sources with well-studied spectra and known fluxes at
several frequencies we determined the expected 3.94-GHz fluxes
${\rm F_{sp}}$ and computed the ratios ${\rm F_{sp}/F_{1}}$, ${\rm
F_{sp}/F_{2}}$ and ${\rm F_{sp}/F_{mean}}$, where  ${\rm F_{1}}$,
${\rm F_{2}}$ и ${\rm F_{mean}}$  are the fluxes of the RCR
sources found using the first and second methods and the average
of the two estimates. Figure~\ref{fig2:Soboleva_n} represents the
corresponding distributions: \ref{fig2:Soboleva_n}a the histogram of the
${\rm F_{sp}/F_{1}}$ ratios; \ref{fig2:Soboleva_n}b the histogram of the
${\rm F_{sp}/F_{2}}$ ratio, and \ref{fig2:Soboleva_n}c the histogram of the
${\rm F_{sp}/F_{mean}}$ ratio, where \mbox{${\rm
F_{mean}=(F_{1}+F_{2})/2}$.}

The median values of the histograms in Fig.~\ref{fig2:Soboleva_n} are equal
to 0.99 (\ref{fig2:Soboleva_n}a), 1.03 (\ref{fig2:Soboleva_n}b), and 0.99 (\ref{fig2:Soboleva_n}c).
The mean $\overline{{\rm F_{sp}/F}}$ ratios averaged over the
entire sample are equal to $0.98\pm0.01$, $1.09\pm0.03$, and
$1.00\pm0.01$ for the fluxes found using the first and second
methods and the averages of the two estimates, respectively.

Hence the the expected 7.6-cm fluxes based on the spectral data
for well-studied radio sources agree best with the halfsum of the
fluxes found using two independent methods. The first and second
methods yielded somewhat over- and underestimated fluxes,
respectively. Recall that the background level is computed with a
80- and 20-s  ``smoothing window'' in the former and latter cases,
respectively.

Thus the use of two different independent methods for extracting
the sources and finding their parameters proved to be efficient
and justified.

We found the following method to be an efficient tool for testing
the reliability of the sources extraction from the scans. All
scans are subdivided into two groups with equal number of scans.
The scans of each group are then averaged and these averaged scans are used to check whether a particular
source is present in the records. The difference between the two scans is then computed and this procedure removes,
with a certain accuracy, all the faint objects that are invisible because of the limited sensitivity (the
saturation effect) and the difference scan shows only the noise record of the receiver and atmospheric noise.

In fig.~\ref{fig16:Soboleva_n} we present the sum (\ref{fig16:Soboleva_n}a) and
difference (\ref{fig16:Soboleva_n}b) of the averaged scans inferred from two
groups of \mbox{11-h} records made in 1994. Weak (about $8\%$)
``negative sources'' indicate the measurements accuracy of the
objects fluxes. We used this technic in the second method of data
reduction.

\section{SPECTRAL INDICES OF THE RCR CATALOG RADIO SOURCES}

In this section we analyze the spectral indices of the radio sources at 3.94~GHz ($\alpha_{3.94}$). To this
end, we subdivide all RCR sources into three groups.

The first group consists of the sources with known fluxes from other catalogs.

The second group consists of bright sources known from other
catalogs and with well-studied spectra with fluxes known at
several frequencies. We used the objects of this group to check
the reliability of the right ascensions and fluxes findings for
objects of the RCR catalog (see the previous section). The sources
of this group are also included into the first group.

The third group consists of the sources with the data available
only from two catalogs---NVSS and RCR---at the  1.4 and 3.94~GHz
frequencies. For some of these sources flux estimates were
obtained based on the GB6 (4.85~GHz) and VLSS (74~MHz) maps.

Let us first consider the second group of sources with well-studied spectra and see how their spectra change
if we supplement the available data with flux estimates based on the GB6 and VLSS maps.

Figure~\ref{fig3:Soboleva_n}a presents the histogram of the spectral indices
of the objects of this group at 3.94~GHz ($\alpha_{3.94}$)
computed with the available data supplemented by the 4.85-GHz flux
estimates (GB6 maps). We found the flux estimates to agree
sufficiently well with the available data including our
observations at 3.94~GHz. If taken into account, the flux
estimates based on GB6 maps change the spectral indices at this
frequency only slightly, on the average by 0.05. This means that
the flux estimates based on GB6 survey maps can also be used to
construct the spectra of the sources for that the data are
available only at two frequencies (i.e., the sources of the third
group). The supplementary data should not change strongly the
spectral indices of the sources considered. At the same time, the
4.85-GHz flux estimates provide indirect check of the reliability
of our 3.94-GHz data for the sources of the third group.

Figure~\ref{fig3:Soboleva_n}b demonstrates the histogram of the spectral
indices of the sources of the second group with well-studied
spectra after improving the spectra with the flux estimates based
on VLSS maps. In a number of cases the use of the 74-MHz flux
estimates had virtually no effect on the computed spectral index
at 3.94~GHz, but in some cases incorporating new data resulted in
an insignificant flattening of the spectrum. However, in most of
the cases the inclusion of flux estimates made the spectrum
steeper at 3.94~GHz and flatter at lower frequencies.

These changes are immediately apparent when we compare
Figs.~\ref{fig3:Soboleva_n}а and~\ref{fig3:Soboleva_n}b: the number of sources with the
spectral indices equal to --0.8 and --1.0 increased in
Fig.~\ref{fig3:Soboleva_n}b and so did the number of sources with the
spectral index \mbox{$\alpha_{3.94} \sim -0.15$.}

In our subsequent computations of the spectra we used the VLSS map
based flux estimates with caution. We used these data primarily in
cases where they were consistent with the data of other catalogs.
In cases where VLSS data were inconsistent with the data of the
TXS catalog we preferred the latter or approximated the spectrum
by a linear relation.

Figure~\ref{fig3:Soboleva_n}c represents the distribution of spectral indices
$\alpha_{3.94}$ for the same group of sources computed with the
flux estimates based on GB6 maps and selected estimates based on
VLSS maps taken into account. We list these indices in the Table.

%Fig.7
\begin{figure}%[tbp]
\onelinecaptionsfalse
\centerline{
\vbox{
\hbox{
\centerline{
\includegraphics[width=8cm]{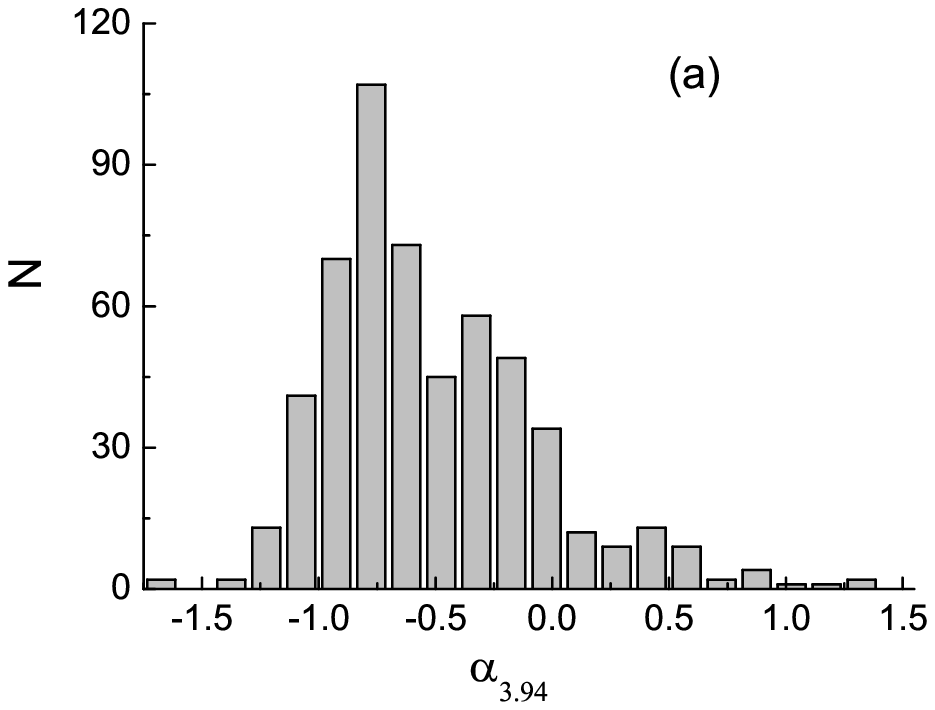}
}} \hbox{ \centerline{
\includegraphics[width=8cm]{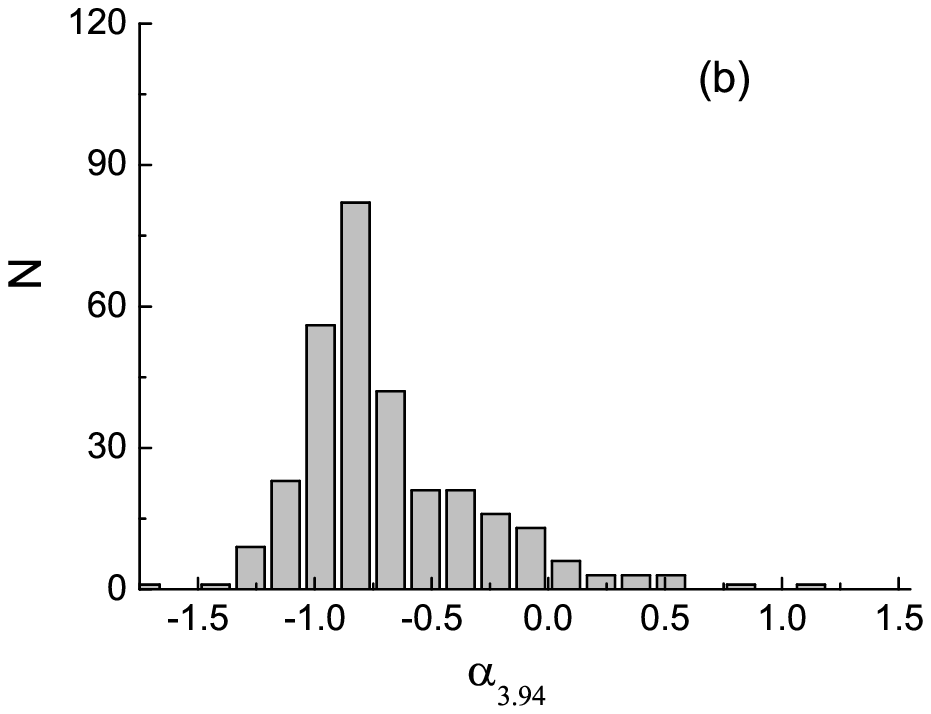}
}} \hbox{ \centerline{
\includegraphics[width=8cm]{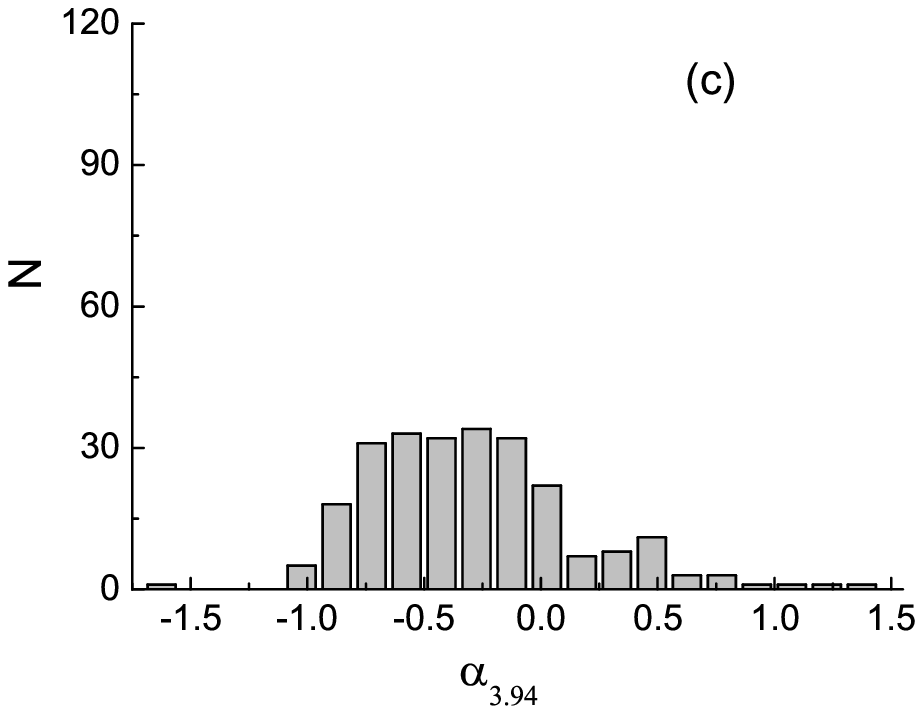}
}} } } \setcaptionmargin{0mm} \captionstyle{normal}
\caption{Histograms of spectral indices at 3.94-GHz: (\ref{fig4:Soboleva_n}a)
for all sources of the RCR catalog; (\ref{fig4:Soboleva_n}b) for the sources
of the first group with flux data available at two or more
frequencies, and \mbox{(\ref{fig4:Soboleva_n}c) } for the sources of the
third group with fluxes known only at two frequencies. }
\label{fig4:Soboleva_n}
\end{figure}

%Fig.8
\begin{figure*}[tbp]
\onelinecaptionsfalse
\centerline{
\hbox{
\centerline{
\includegraphics[angle=0,width=0.45\textwidth,clip]{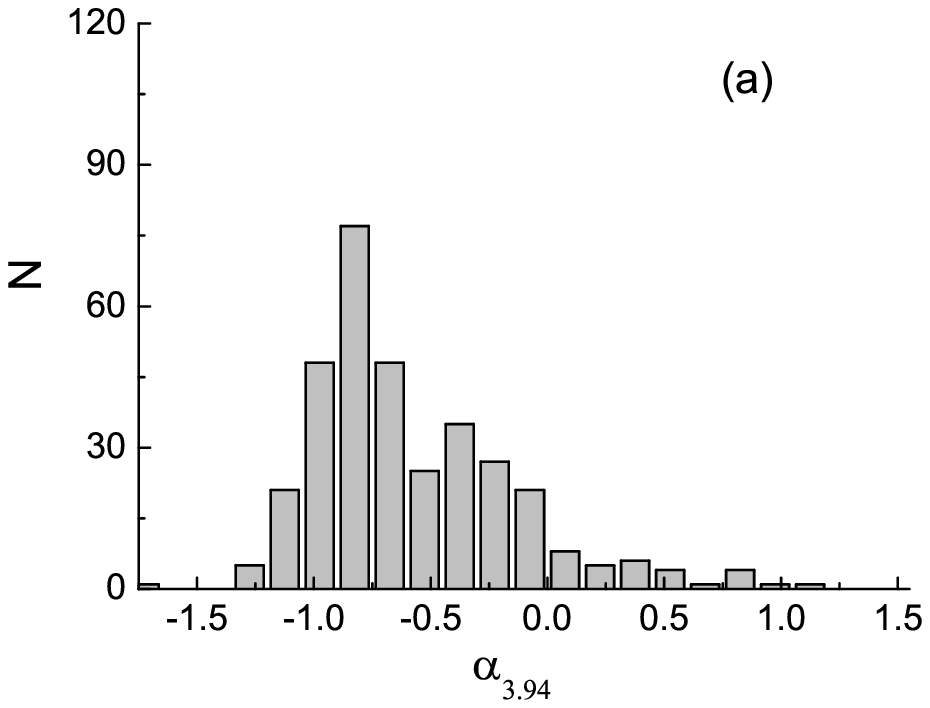}
\includegraphics[angle=0,width=0.45\textwidth,clip]{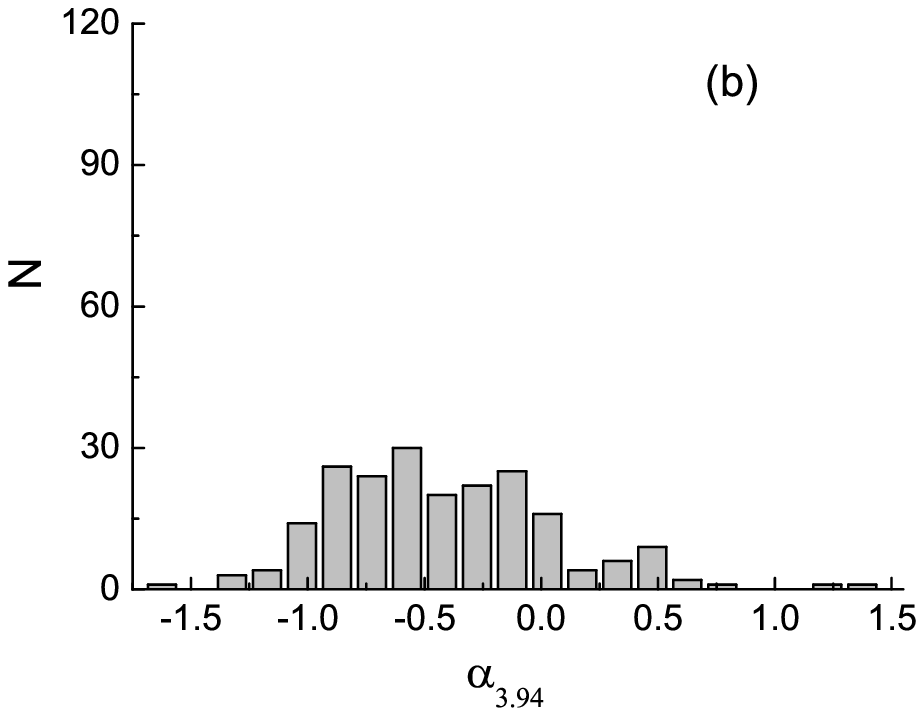}
} } } \setcaptionmargin{0mm} \captionstyle{normal}
\caption{Histograms of spectral indices at 3.94 GHz for sources
with antenna temperatures ${\rm T_{a}} > 5 \sigma$ (\ref{fig4a:Soboleva_n}a)
and $3 \sigma \le  {\rm T_{a}} < 5 \sigma$  (\ref{fig4a:Soboleva_n}b). }
\label{fig4a:Soboleva_n}
\end{figure*}

%Fig.9
\begin{figure*}[tbp]
\onelinecaptionsfalse
\centerline{
\vbox{
\hbox{
\centerline{
\includegraphics[angle=0,width=0.45\textwidth,clip]{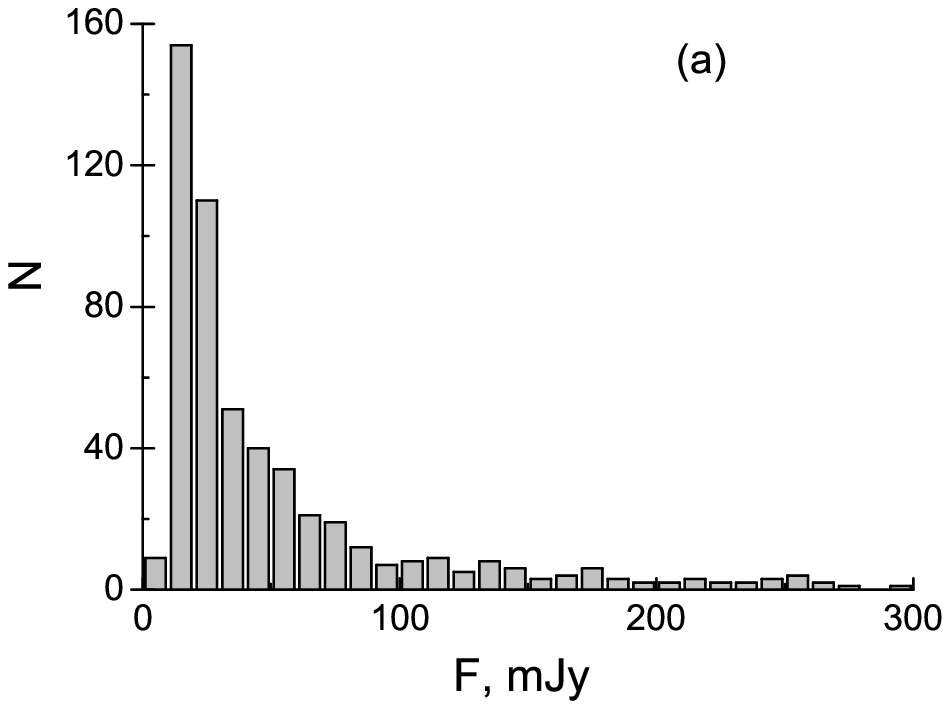}
\includegraphics[angle=0,width=0.45\textwidth,clip]{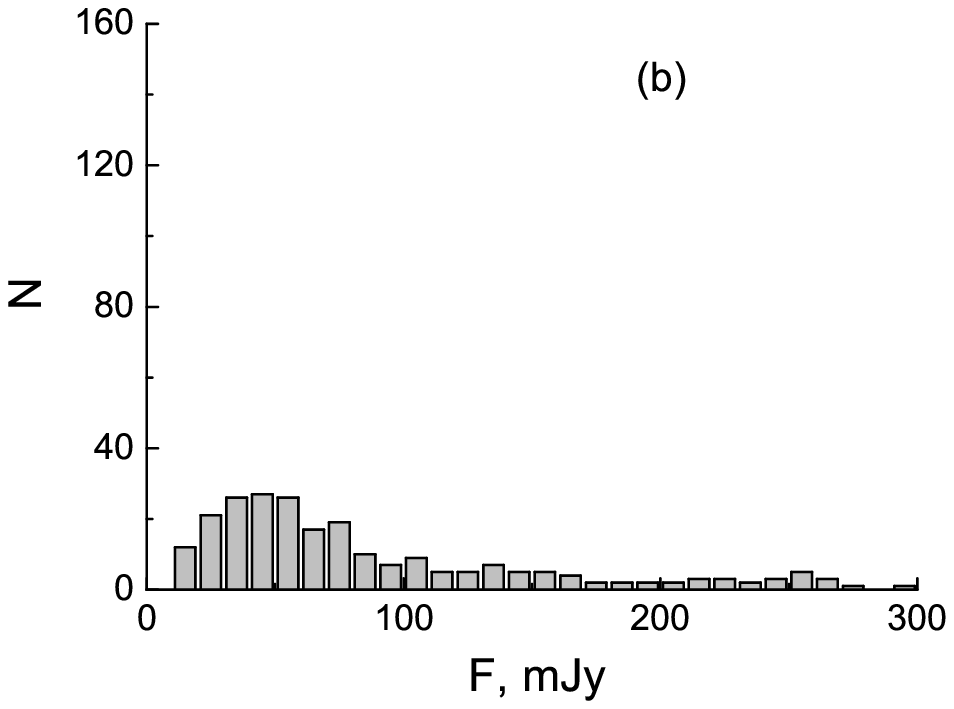}
}
}
\hbox{
\centerline{
\includegraphics[angle=0,width=0.45\textwidth,clip]{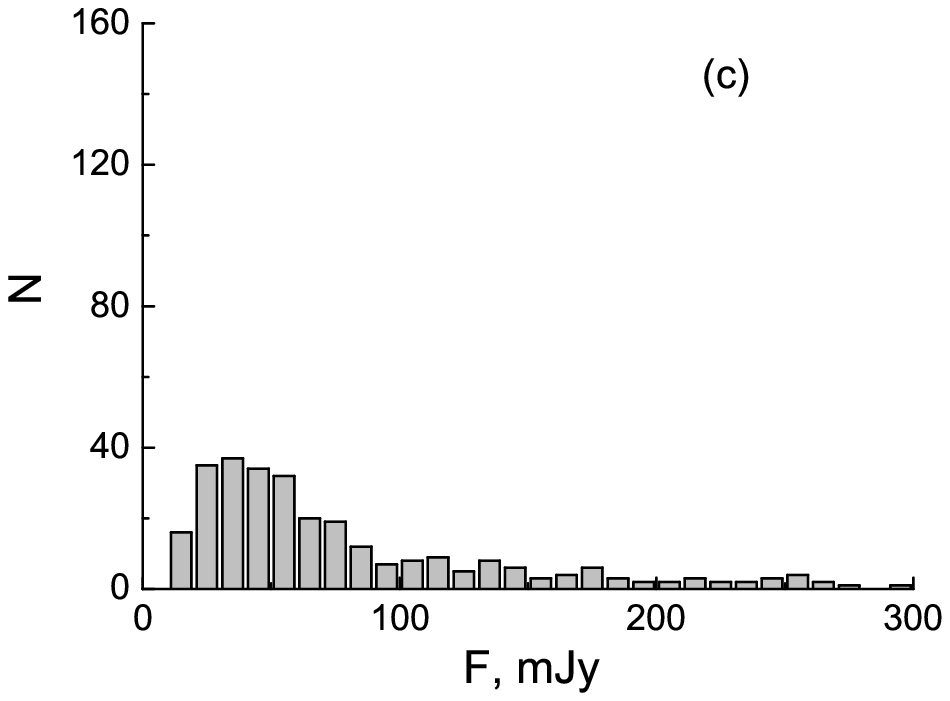}
\includegraphics[angle=0,width=0.45\textwidth,clip]{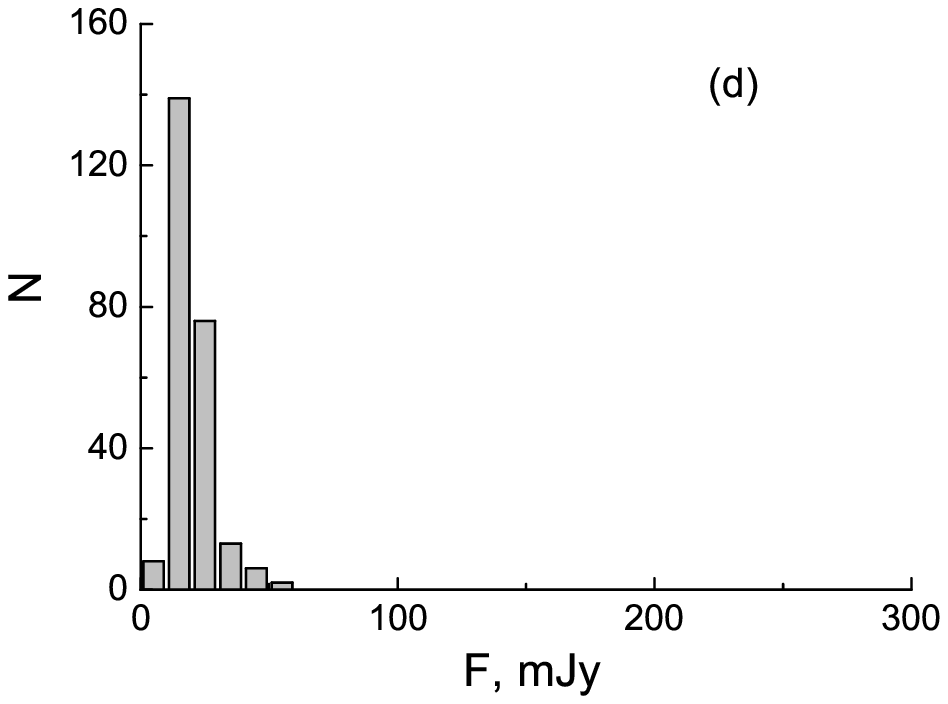}
} } } } \setcaptionmargin{0mm} \captionstyle{normal}
\caption{Histograms of 3.94-GHz fluxes for all sources of the RCR
catalog (\ref{fig5:Soboleva_n}a), for sources of the second (\ref{fig5:Soboleva_n}b),
first (\ref{fig5:Soboleva_n}c), and third (\ref{fig5:Soboleva_n}d) groups.} \label{fig5:Soboleva_n}
\end{figure*}

%Fig.10
\begin{figure*}[tbp]
\onelinecaptionsfalse
\centerline{
\hbox{
\centerline{
\includegraphics[angle=0,width=0.45\textwidth,clip]{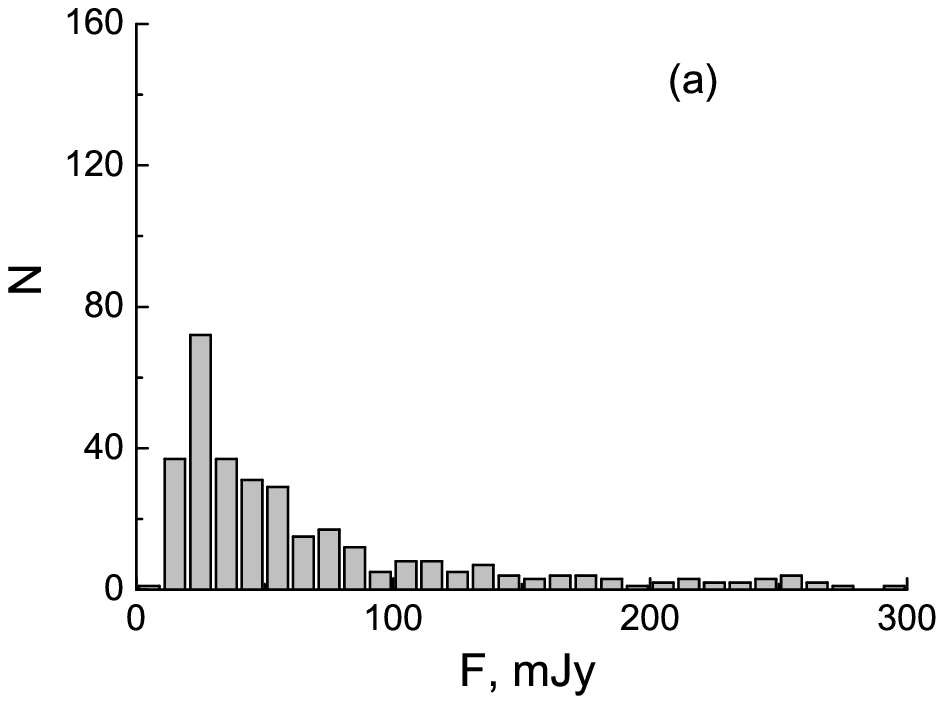}
\includegraphics[angle=0,width=0.45\textwidth,clip]{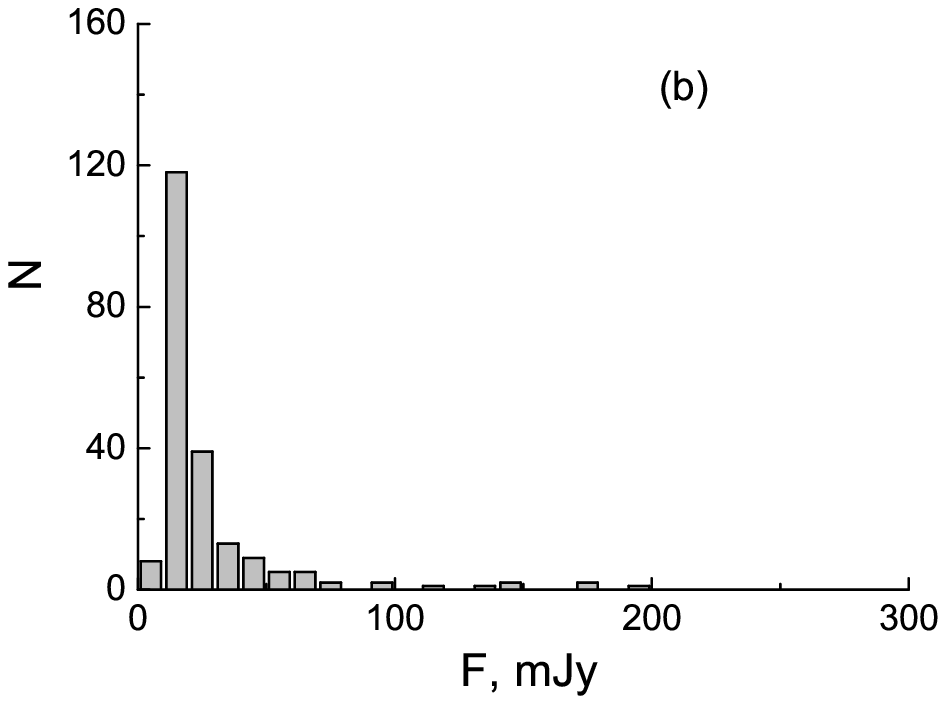}
} } } \setcaptionmargin{0mm} \captionstyle{normal}
\caption{Histograms of the 3.94-GHz fluxes for sources with
antenna temperatures ${\rm T_{a}} > 5 \sigma$ (\ref{fig5a:Soboleva_n}a) and
$3 \sigma \le  {\rm T_{a}} < 5 \sigma$ (\ref{fig5a:Soboleva_n}b).}
\label{fig5a:Soboleva_n}
\end{figure*}

%Fig.11
\begin{figure}[tbp]
\onelinecaptionsfalse
\centerline{
\hbox{
\includegraphics[angle=0,width=0.5\textwidth,clip]{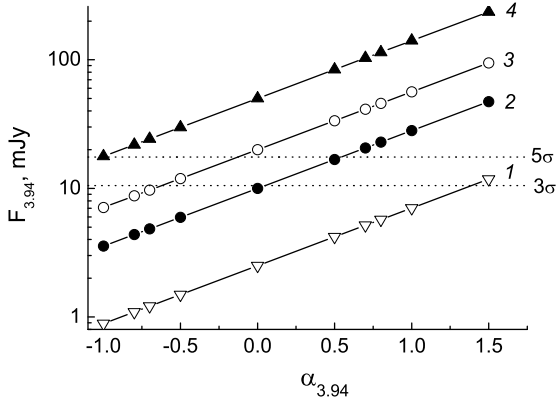}
} } \setcaptionmargin{0mm} \captionstyle{normal} \caption{Expected
7.6-cm fluxes for different 1.4-GHz fluxes (NVSS objects) as a
function of spectral index. The dotted horizontal lines indicate
the threshold of object detection at the  $3\sigma$ and $5\sigma$
levels for the sensitivity achieved at RATAN-600 radio telescope.
The straight lines {\it 1}, {\it 2}, {\it 3}, and {\it 4}
correspond to NVSS objects with fluxes ${\rm F_{1.4}} > 2.5$ mJy,
${\rm F_{1.4}} > 10$ mJy, ${\rm F_{1.4}} > 20$ mJy, and ${\rm
F_{1.4}} > 50$ mJy, respectively. The dependences are given for
the central section of the survey. } \label{fig55:Soboleva_n}
\end{figure}

%Fig.12
\begin{figure}[tbp]
\onelinecaptionsfalse
\centerline{
\hbox{
\includegraphics[angle=0,width=0.5\textwidth,clip]{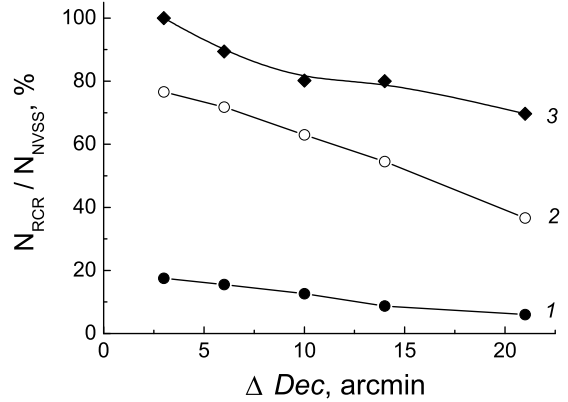}
} } \setcaptionmargin{0mm} \captionstyle{normal} \caption{ Ratio
of the number of RCR objects in the given declination band
$Dec_{0} \pm \Delta Dec$ to the number of NVSS sources in the same
band as a function of the band width  $\Delta Dec$ for several
intervals of NVSS object fluxes: ${\rm F_{1.4}} < 20$  mJy (curve
{\it 1}), \mbox{20 mJy$  \le  {\rm F_{1.4}} < 100$  mJy} (curve
{\it 2}), and $ {\rm F_{1.4}} > 100$  mJy \mbox{(curve {\it 3})}.
} \label{fig15:Soboleva_n}
\end{figure}

%Fig.13
\begin{figure}[tbp]
\onelinecaptionsfalse
\centerline{
\vbox{
\hbox{
\centerline{
\includegraphics[angle=0,width=0.45\textwidth,clip]{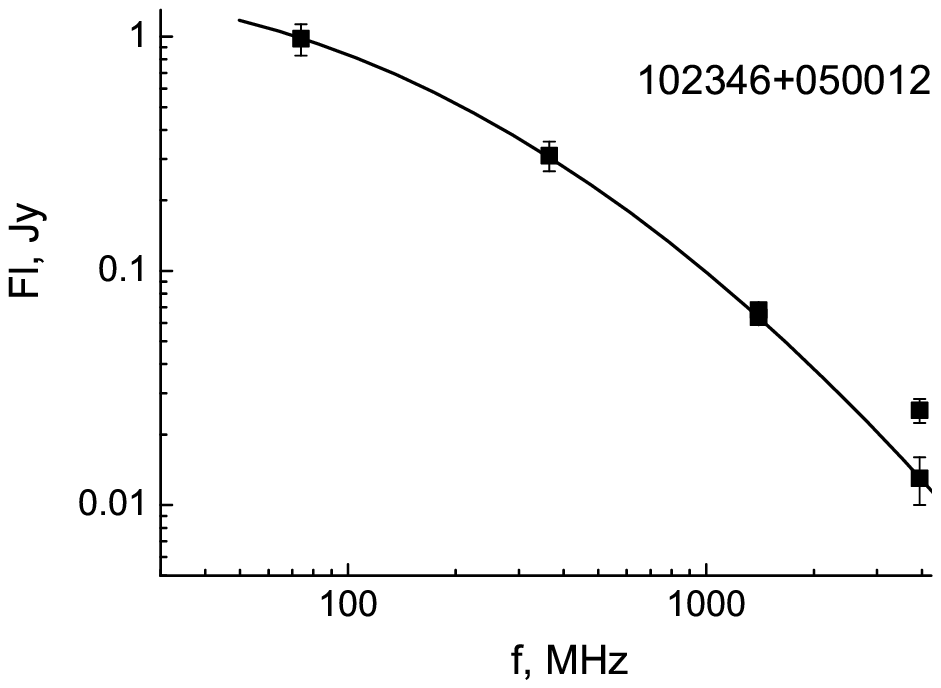}
}}
\hbox{
\centerline{
\includegraphics[angle=0,width=0.45\textwidth,clip]{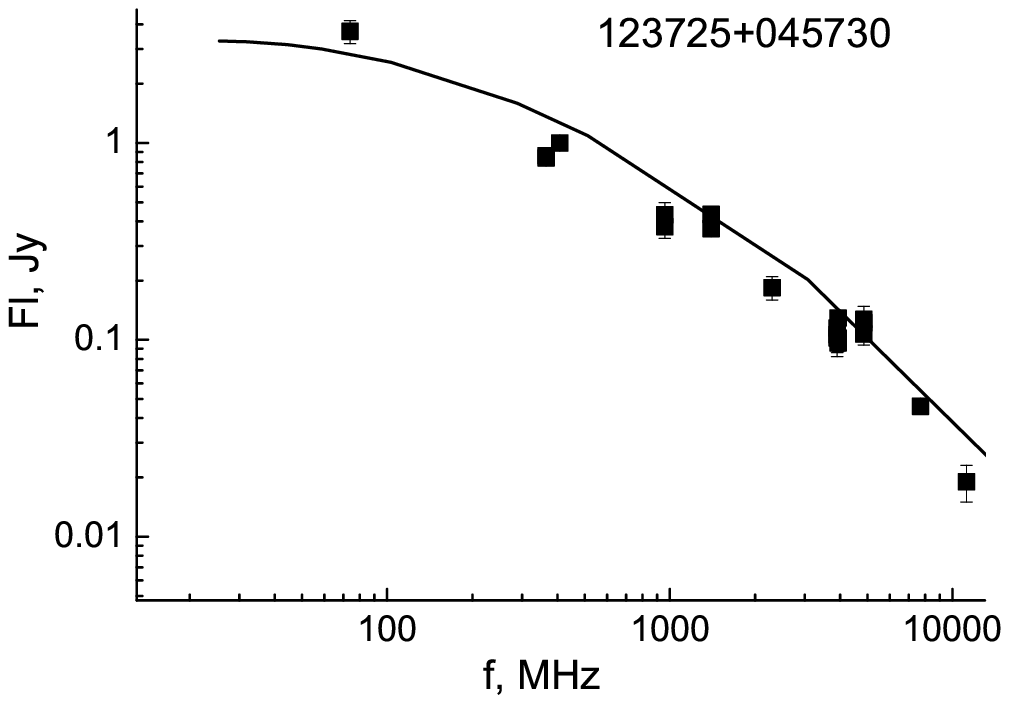}
}}
\hbox{
\centerline{
\includegraphics[angle=0,width=0.45\textwidth,clip]{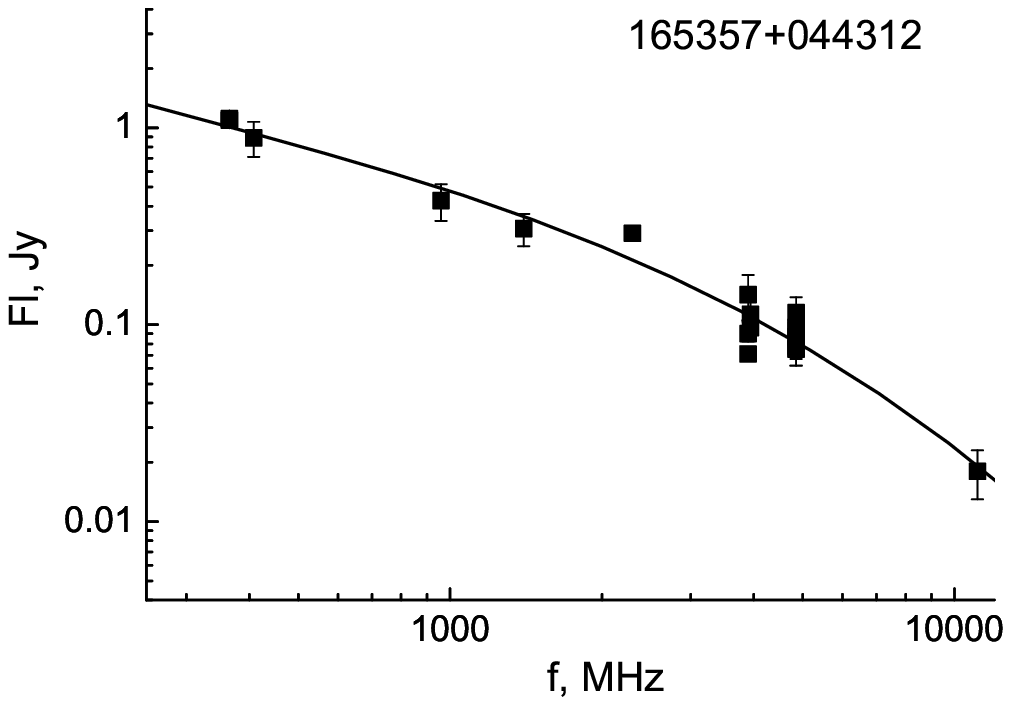}
}
}}}
\setcaptionmargin{0mm}
\captionstyle{normal}
\caption{Examples of spectra with the steepness increasing at high frequencies.
}
\label{fig7:Soboleva_n}
\end{figure}

%Fig.14
\begin{figure*}[tbp]
\onelinecaptionstrue
\centerline{
\hbox{
\includegraphics[angle=0,width=0.45\textwidth,clip]{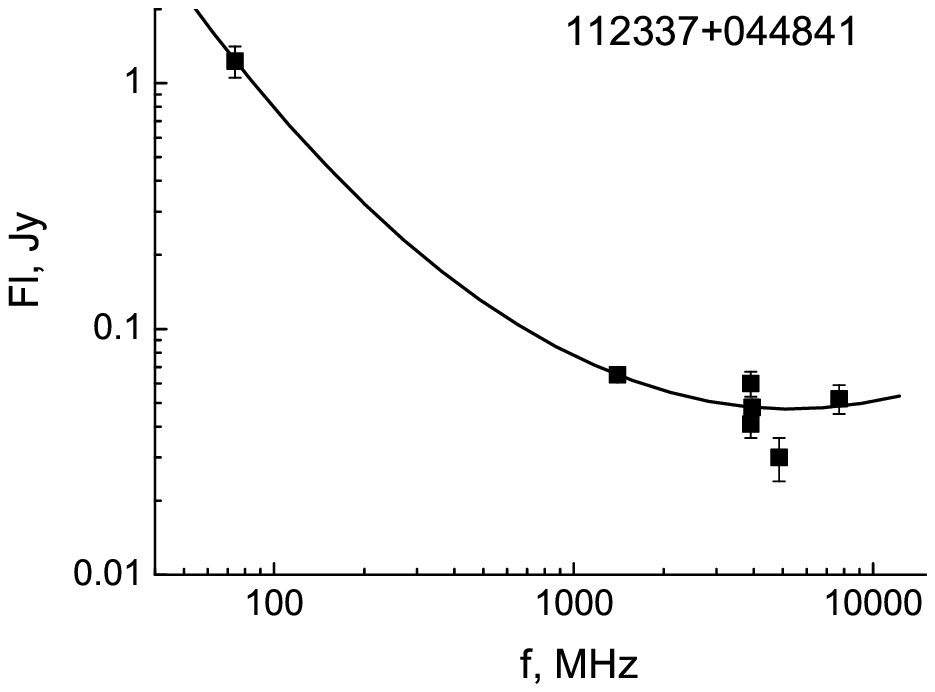}
\includegraphics[angle=0,width=0.45\textwidth,clip]{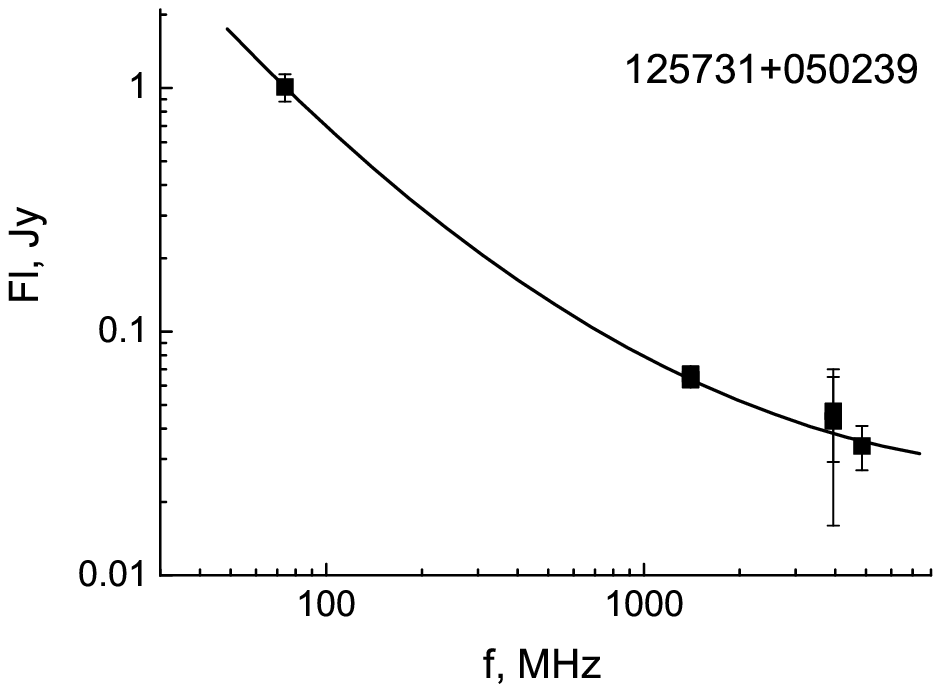}
}
}
\setcaptionmargin{0mm}
\captionstyle{normal}
\caption{Examples of spectra with the steepness increasing at low frequencies.
}
\label{fig8:Soboleva_n}
\end{figure*}

%Fig.15
\begin{figure}[tbp]
\onelinecaptionsfalse
\centerline{
\hbox{
\includegraphics[angle=0, width=0.5\textwidth,clip]{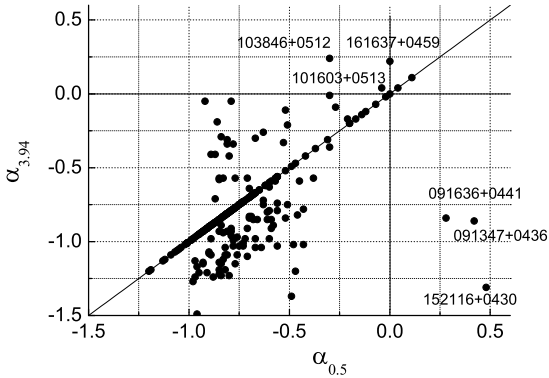}
}
}
\setcaptionmargin{0mm}
\captionstyle{normal}
\caption{Spectral indices $\alpha_{3.94}$ plotted versus $\alpha_{0.5}$ for the radio sources of the catalog.
}
\label{fig10:Soboleva_n}
\end{figure}

%Fig. 16
\begin{figure*}[tbp]
\onelinecaptionstrue
\centerline{
\vbox{
\hbox{
\centerline{
\includegraphics[angle=0,width=0.45\textwidth,clip]{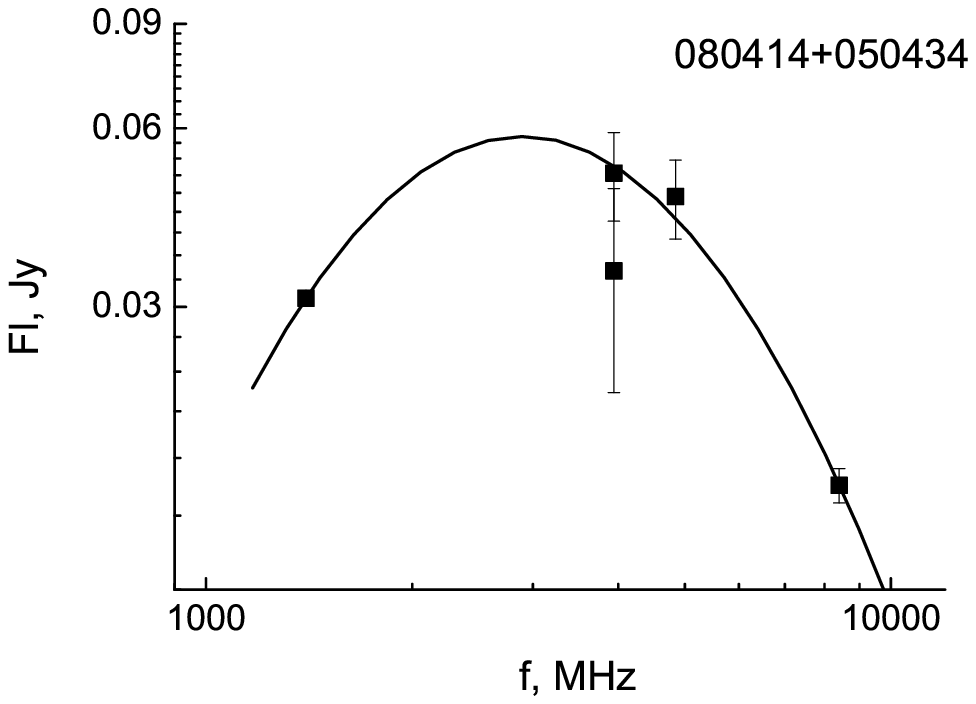}
\includegraphics[angle=0,width=0.45\textwidth,clip]{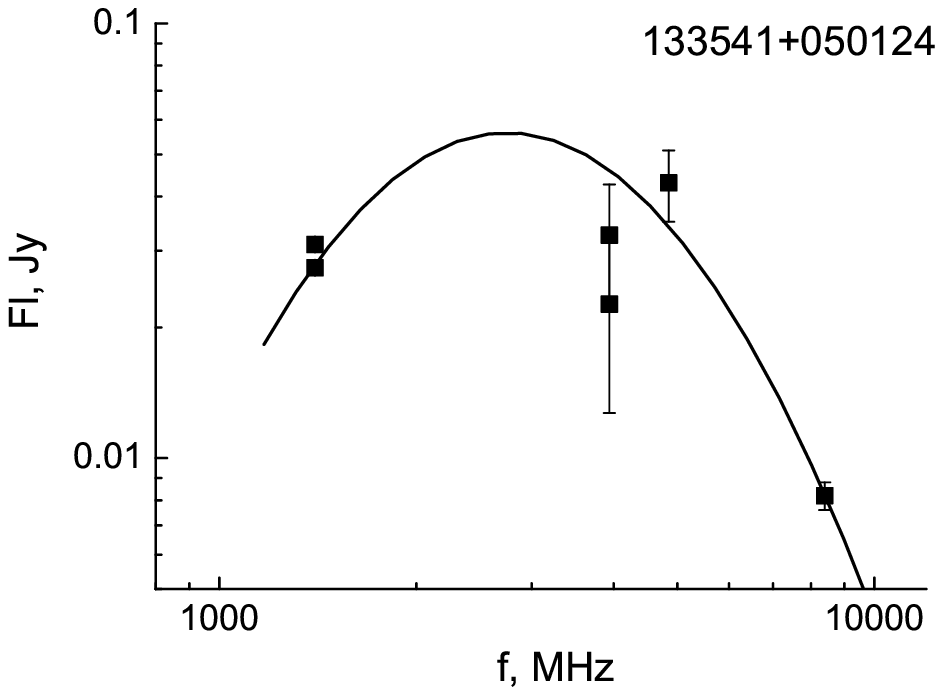}
}}
\hbox{
\centerline{
\includegraphics[angle=0,width=0.45\textwidth,clip]{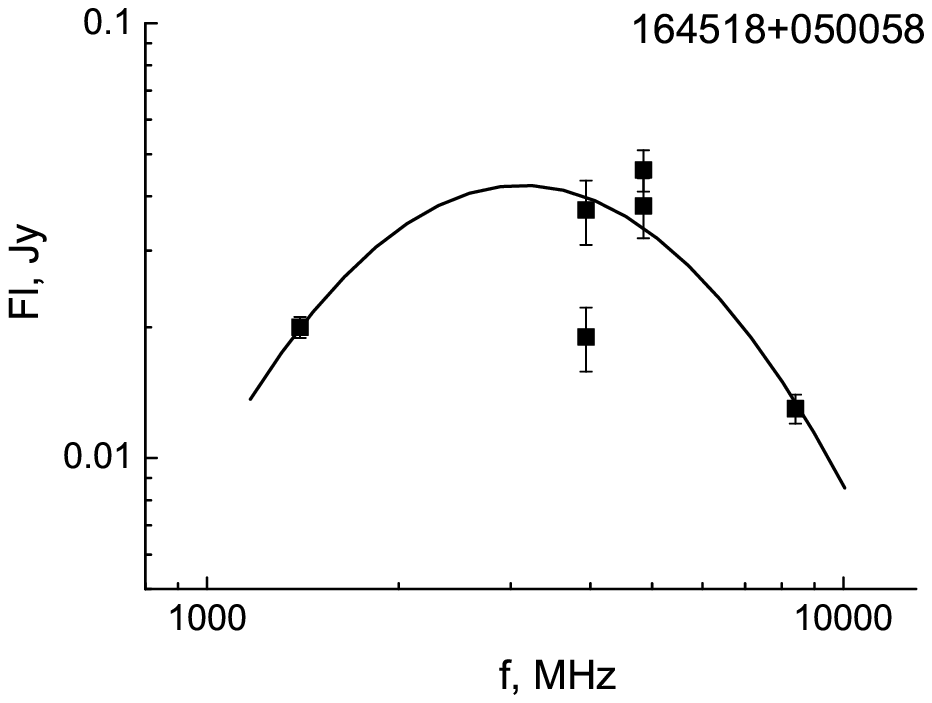}
\includegraphics[angle=0,width=0.45\textwidth,clip]{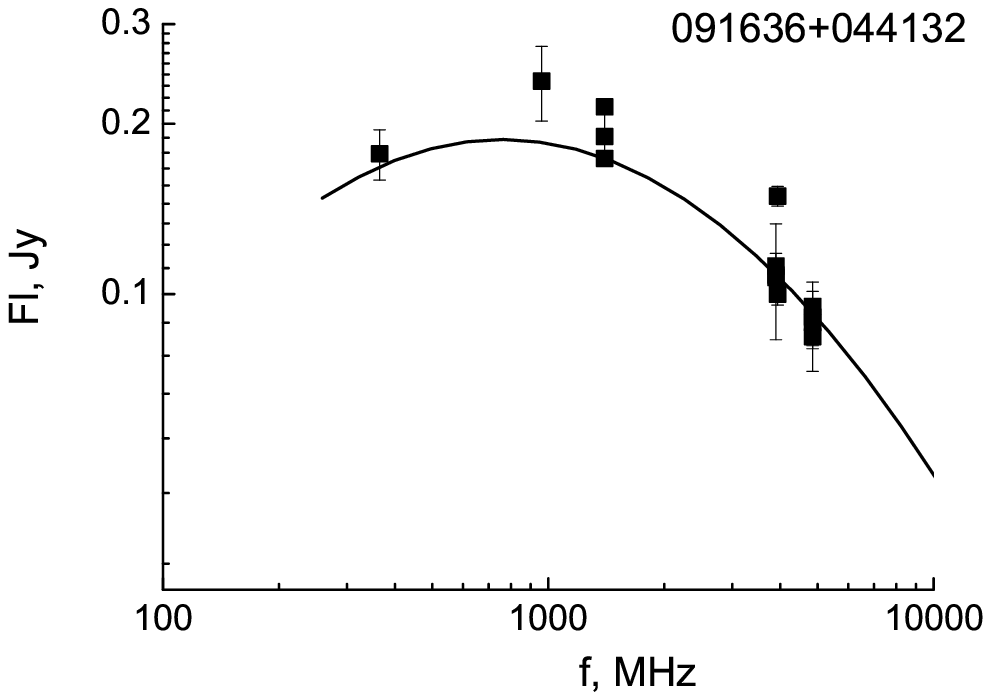}
}
}
\hbox{
\centerline{
\includegraphics[angle=0,width=0.45\textwidth,clip]{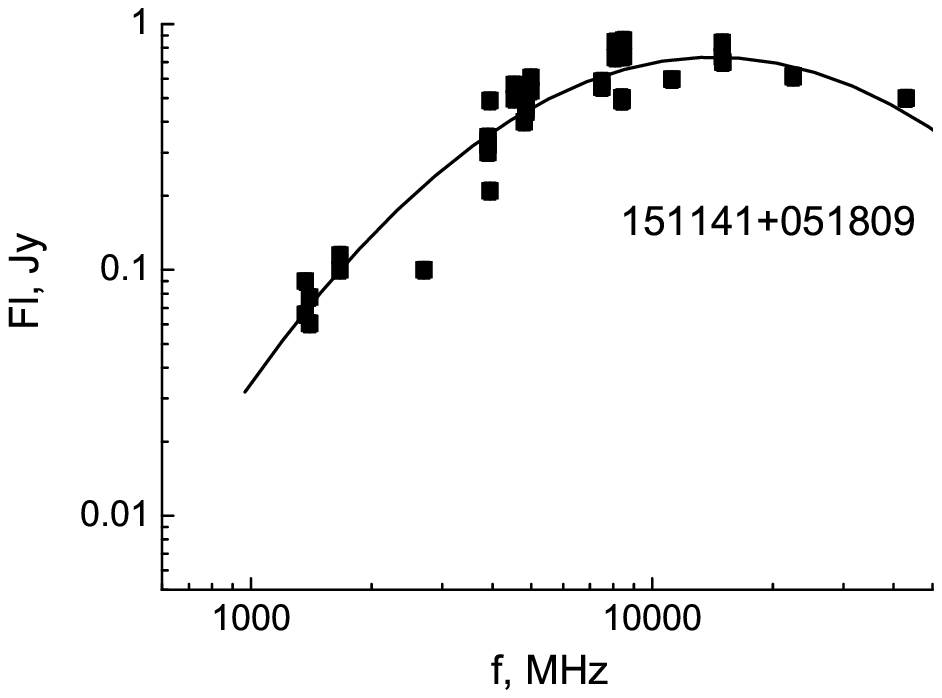}
}
}
}
}
\setcaptionmargin{0mm}
\captionstyle{normal}
\caption{
Examples of spectra with a maximum at some frequency. In the Catalog these sources are marked as GPS
(column~8).
}
\label{fig6:Soboleva_n}
\end{figure*}

%Fig.17
\begin{figure}[tbp]
\onelinecaptionsfalse
\centerline{
\vbox{
\hbox{
\centerline{
\includegraphics[width=8cm]{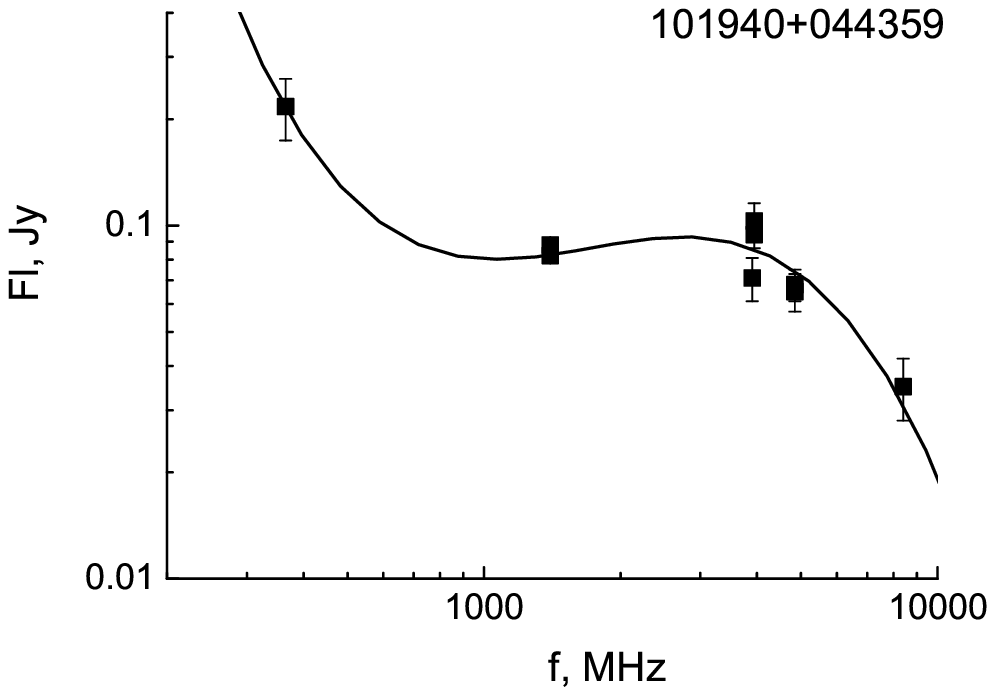}
}} \hbox{ \centerline{
\includegraphics[width=8cm]{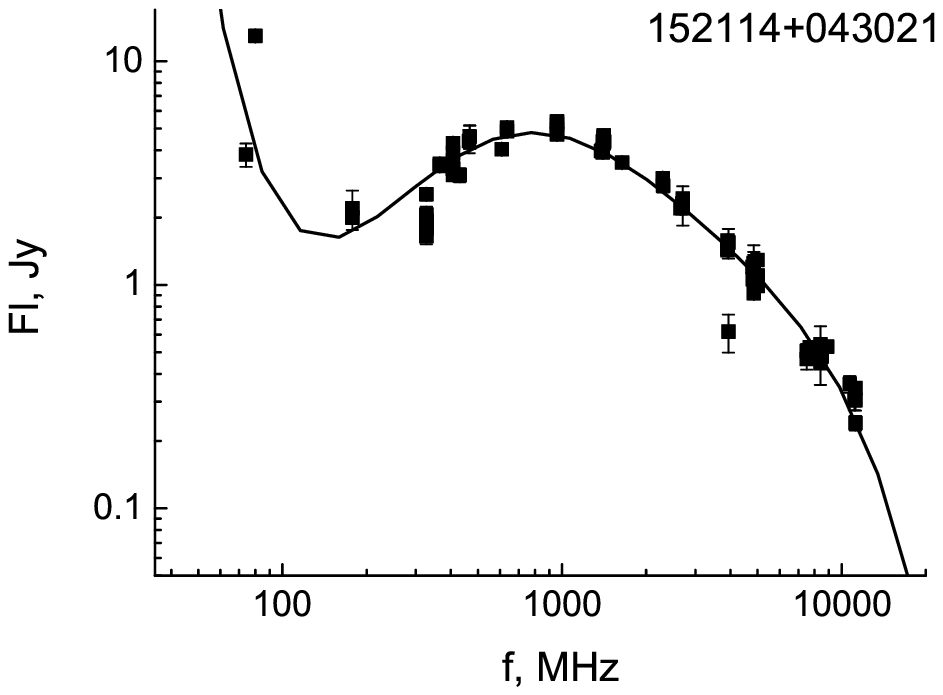}
}} \hbox{ \centerline{
\includegraphics[width=8cm]{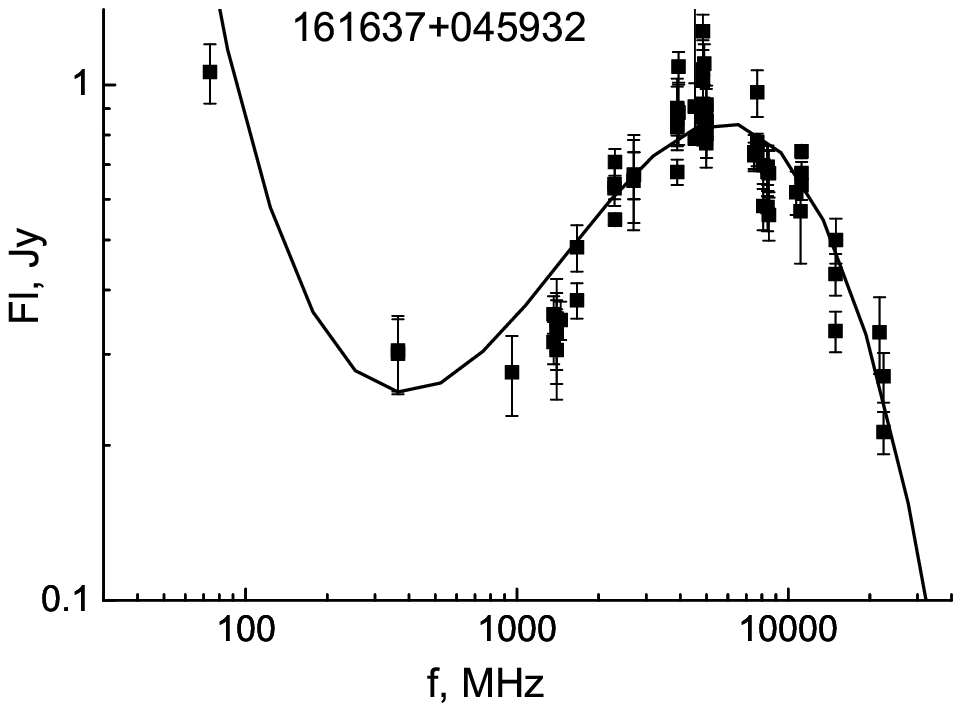}
} }}} \setcaptionmargin{0mm} \captionstyle{normal}
\caption{Examples of spectra resulting from a superposition of a
power-law spectrum with a self-absorption spectrum at frequencies
from 0.5 to 12~GHz. In the Catalog objects with the
101940+044359-source-type spectra are marked as ``hill''
(column~8) (the upper panel). } \label{fig11:Soboleva_n}
\end{figure}

Figure~\ref{fig4:Soboleva_n}a gives us the histogram of the spectral indices
$\alpha_{3.94}$ for all objects of the RCR catalog located in the
studied right ascension interval \mbox{$7^h \le {\rm R.A.} <
17^h$}. Figures~\ref{fig4:Soboleva_n}b and \ref{fig4:Soboleva_n}c show the histograms of
the spectral indices for objects of the first and third groups,
respectively.

In addition, we also constructed the histograms of the spectral
indices of objects with the antenna temperatures on the averaged
scans obeying the conditions ${\rm T_{a}} \ge  5 \sigma$
(Fig.~\ref{fig4a:Soboleva_n}a) and 3$\sigma \le {\rm T_{a}} < 5 \sigma$
(Fig.~\ref{fig4a:Soboleva_n}b). In the Table the latter sources are marked by
asterisks.

It follows from the histograms shown in the figures that most of the objects of the RCR catalog with well-known
fluxes at many frequencies have standard power-law spectra with spectral indices in the interval
$  -1.12   < \alpha_{3.94} < -0.5$.

The distributions of spectral indices in Fig.~\ref{fig4:Soboleva_n}a exhibit
three well-defined maxima and those shown in Figs.~\ref{fig3:Soboleva_n},
\ref{fig4:Soboleva_n}b, and \ref{fig4a:Soboleva_n}a---two maxima, that are indicative of
the presence of two or more radio sources populations among the
objects of our list. The maxima in Fig.~\ref{fig4:Soboleva_n}a correspond to
the spectral indices of \mbox{${\alpha_{3.94}}_{max} = -0.8,
-0.3$}, and +0.45.

The second maximum shows up less clearly for objects of the third
group for that fluxes are known only at two frequencies. The main
maximum is at ${\alpha_{3.94}}_{max} \sim -0.45$ and the less
conspicuous maximum is at ${\alpha_{3.94}}_{max} \sim +0.50$.

Objects with antenna temperatures ${\rm T_{a}} \ge  5 \sigma$
\linebreak (Fig.~\ref{fig4a:Soboleva_n}a) exhibit a more conspicuous second
maximum in the distribution of spectral indices than the objects
of the first group. Their distribution is more like that of the
objects of the second group---i.e., objects with well-studied
spectra.

%Fig.18
\begin{figure}[tbp]
\onelinecaptionsfalse
\centerline{
\vbox{
\centerline{
\hbox{
\includegraphics[angle=0,width=0.46\textwidth,clip]{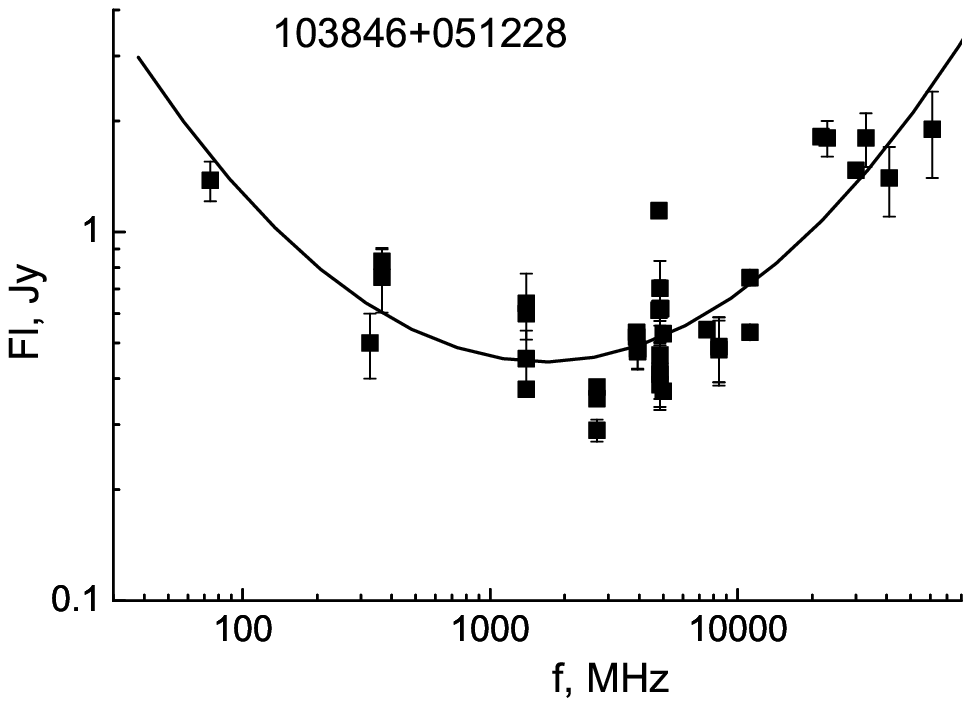}
}}
\centerline{
\hbox{
\includegraphics[angle=0,width=0.46\textwidth,clip]{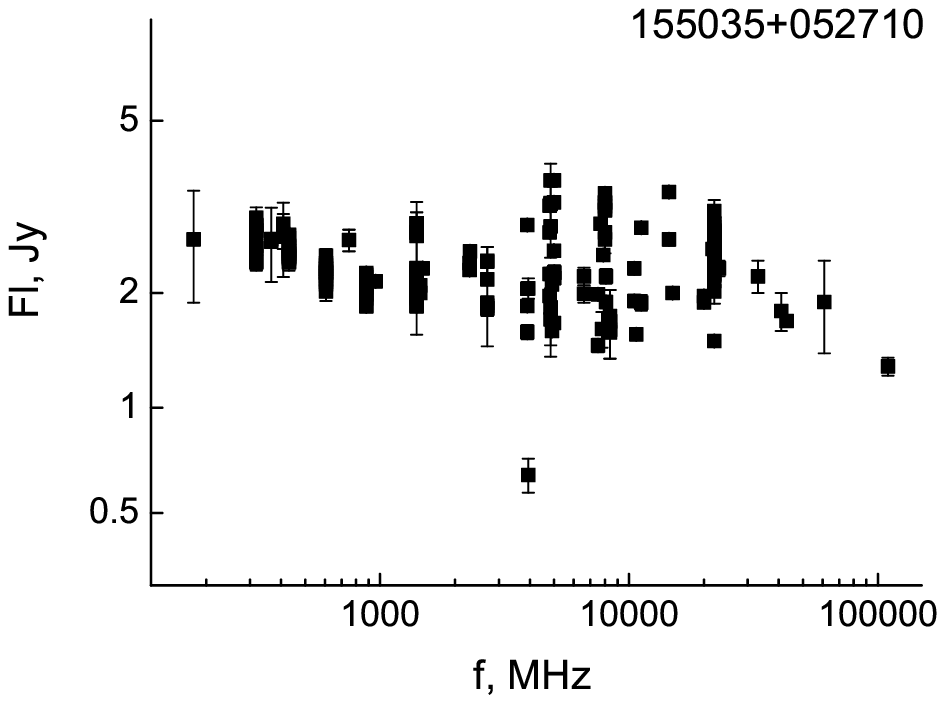}
}
}}}
\setcaptionmargin{0mm}
\captionstyle{normal}
\caption{Spectrum with the minimum at a certain frequency (the upper panel) and the spectrum of the variable
source (the lower panel).
}
\label{fig12:Soboleva_n}
\end{figure}

There is also a considerable similarity between the distributions
of the spectral indices of objects with antenna temperatures
3$\sigma \le {\rm T_{a}} < 5 \sigma$ and objects of the third
group, for that the spectral data are known only at two
frequencies. Such a similarity is quite natural. The lower the
sensitivity threshold (3$\sigma$) is, the fainter sources we
extract from the records. However, in this case we also face a
higher risk of extracting artifacts.

The sample of sources that we extracted at the 3$\sigma$--$5\sigma$ level intersects only partially
with the sample of objects of the third group. It contains more objects with steep spectra
($\alpha_{3.94} < -0.75$) than the sample of objects of the third group, whereas the spectral distributions
are almost the same in the  $\alpha_{3.94} \ge 0$ interval.

Figure~\ref{fig5:Soboleva_n} presents the histograms of the 3.94-GHz fluxes
for the entire list of the RCR catalog (\ref{fig5:Soboleva_n}a), the sources
of the second (\ref{fig5:Soboleva_n}b), first (\ref{fig5:Soboleva_n}с), and third
(\ref{fig5:Soboleva_n}d) groups.

Figure~\ref{fig5a:Soboleva_n} demonstrates the histograms of the fluxes of
objects with the antenna temperatures \mbox{$ {\rm T_{a}} \ge  5
\sigma$} on the averaged scans (\ref{fig5a:Soboleva_n}a) and for sources with
${\rm T_{a}}$ that satisfy the condition 3$\sigma \le {\rm T_{a}}
< 5 \sigma$ (\ref{fig5a:Soboleva_n}b).

The fluxes of the first and second groups sources and the sources
with the antenna temperatures \mbox{${\rm T_{a}}  \ge  5 \sigma$}
lie in the interval from 10~mJy to several Jy (the x-axis domains
in the figures are limited by  300~mJy).

The fluxes of sources with the antenna temperatures  3$\sigma \le
{\rm T_{a}} < 5 \sigma$ also exhibit a large scatter, from 10 to
200~mJy. The distribution of the fluxes of this sample objects
resembles that of the third group objects. The difference is in
that all objects of the third group are faint, with the 3.94-GHz
fluxes of 60~mJy or less (Fig.~\ref{fig5:Soboleva_n}d). About 95\% of them
have fluxes that do not exceed $30$~mJy, and most of the sources
(about $85\%$) have fluxes in the 10--20\,mJy interval.

Number of faint objects with flat or inverse spectra is easy to
explain by examining Fig.~\ref{fig55:Soboleva_n}, that demonstrate the
changes in the expected 7.6-cm fluxes as a function of spectral
index for objects with different 1.4-GHz fluxes  (NVSS objects).

The horizontal dashed lines in the figure indicate the possibility
of objects detection at the $3\sigma$ and  $5\sigma$ levels for
the sensitivity achieved at \mbox{RATAN-600}. The straight lines
{\it 1}, {\it 2}, {\it 3}, and {\it 4} correspond to NVSS objects
with fluxes  ${\rm F_{1.4}} > 2.5$~mJy, ${\rm F_{1.4}} > 10$~mJy,
${\rm F_{1.4}} > 20$~mJy, and ${\rm F_{1.4}} > 50$~mJy,
respectively. These dependences are for the central section of the
survey.

It follows from Fig.~\ref{fig55:Soboleva_n} that at the $3\sigma$
level \linebreak \mbox{RATAN-600} could have detected at 7.6~cm
the faintest NVSS objects (2.5~mJy $<  {\rm F_{1.4}} < 10$ mJy)
with inverse spectral indices \mbox{$\alpha > 1.5$}  (the
population of SSA (Synchrotron Self Absorption) and blackbody
spectrum objects).

The $3\sigma$ level in the central band is close to 10~mJy, and,
starting with a certain integration time, it is virtually
independent of further integration due to the saturation effect.
We can therefore expect that at this detection level all NVSS
objects with fluxes \mbox{${\rm F_{1.4}} > 20$~mJy} and with
spectral indices $\alpha > 0$ crossing the central section of our
survey can be detected and must have been occurred in the RCR
catalog.

We determined the completeness of our catalog from the ratio of
the number of objects of the RCR catalog in the given declination
interval \mbox{$Dec_{0} \pm \Delta Dec$} to the number of sources
of the NVSS catalog in the same band for several flux intervals of
NVSS objects: $ {\rm F_{1.4}} < 20$\,mJy (curve {\it 1} in
Fig.~\ref{fig15:Soboleva_n}), \mbox{20\,mJy$  \le  {\rm F_{1.4}} \le
100$\,mJy} (curve {\it 2} in Fig.~\ref{fig15:Soboleva_n}), and $ {\rm
F_{1.4}} > 100$~mJy (curve 3 in Fig.~\ref{fig15:Soboleva_n}).

It follows from Fig.~\ref{fig15:Soboleva_n} that the completeness of our
catalog increases with increasing flux of NVSS objects and with
decreasing survey band. Thus in the  $\Delta Dec = \pm6'$ band a
total of 90$\%$ RCR sources have been identified with NVSS objects
with fluxes ${\rm F_{1.4}} \ge  100 $\,mJy, 72$\%$ are identified
with  NVSS objects with fluxes  20~mJy  < ${\rm F_{1.4}}$ <
100\,mJy, and 16$\%$ are identified with NVSS objects with
\mbox{${\rm F_{1.4}} \le  20$\,mJy.} Increasing the survey band to
$\Delta Dec = \pm 20'$ reduces the number of sources identified
with NVSS objects down to  70$\%$ (${\rm F_{1.4}} \ge $ 100~mJy),
37$\%$ \linebreak \mbox{(20\,mJy < ${\rm F_{1.4}}$ <100\,mJy)},
and 6$\%$ (${\rm F_{1.4}} \le $ 20\,mJy), respectively.

\section{SPECTRA OF THE RCR CATALOG RADIO SOURCES}

We make it clear from the start that we cannot report the spectra
of all radio sources of our list in this paper. We will publish
them in a special report and at the  {\tt
http://www.sao.ru/hq/len} web page. Here we classify the spectra
obtained and discuss the most important spectra. Below we present
examples of spectra.

Some of the spectra exhibit a sharp maximum in the frequency interval considered. In our Catalog we mark the sources
with such spectra as  hill, HFP, or GPS (column~8).

Most of the NVSS objects in the RCR catalog with fluxes known at many frequencies have standard power-law spectra
with spectral indices in the interval \mbox{$-1.12  < \alpha_{3.94} < -0.5$.} These spectra can be fitted quite
well by linear relations.

About $20\%$ of the sources have power-law spectra that become
steeper at high frequencies (Fig.~\ref{fig7:Soboleva_n}). Such spectra are
usually interpreted as a result of energy loss by energetic
relativistic electrons via radiative cooling.

About $10\%$ of objects have spectra that become flatter at high
frequencies (Fig.~\ref{fig8:Soboleva_n}). For most of these objects fluxes
are known only at three frequencies: 0.74\,GHz (VLSS), 1.4\,GHz
(NVSS), and 3.94\,GHz (RCR); 0.74\,GHz fluxes are often just
estimates based on the maps of the VLSS survey. The spectra of
such sources are believed to be a result of the superposition of a
common power-law spectrum from extended components of the radio
source with the spectrum from small angular size features (jets
emerging from the nucleus) with synchrotron self absorption by
relativistic electrons at higher frequencies. In principle,
extended components may have different spectra (e.g., steep plus
flat), however, they usually have similar spectra and this is
unlikely.

We pointed out above that most of the objects have power-law
spectra with $\alpha_{3.94}/\alpha_{0.5} = 1$. In
Fig.~\ref{fig10:Soboleva_n}, where we plot spectral indices $\alpha_{3.94}$
versus $\alpha_{0.5}$, these sources lie along the
$\alpha_{3.94}/\alpha_{0.5}$ line. There are a total of about 160
sources with linear spectra; about 70 sources with spectra that
become steeper at high frequencies, and about 25 sources with
spectra that become steeper at low frequencies. In the domain of
negative spectral indices in Fig.~\ref{fig10:Soboleva_n} the former and
latter lie below and above the $\alpha_{3.94}/\alpha_{0.5}$ line,
respectively.

The fourth type of spectra are those with a well-defined maximum
in the frequency interval from 0.5 to $12$\,GHz. We present
examples of such spectra in Figs.~\ref{fig6:Soboleva_n} and \ref{fig11:Soboleva_n}. In
the literature compact sources with maxima in the spectra are
subdivided into three groups: CSS (Compact Steep Spectrum) objects
with the maxima at frequencies lower 0.5~GHz; GPS (Gigahertz Peak
Spectrum) sources with the maxima at frequencies between 0.5 and
5~GHz, and HFP (High Frequency Spectrum) sources with the maxima
at frequencies higher 5~GHz. Sources with a maximum in the
spectrum are believed to be either young objects or blazars
\cite{ode:Soboleva_n,ti1:Soboleva_n,ti2:Soboleva_n}.

The RCR catalog contains 19 objects with the spectra of the fourth type. A minor part
of these objects (six sources) have spectra in the form of the superposition of a power-law
spectrum at low frequencies onto a self-absorption spectrum at frequencies from 0.5 to
12\,GHz.

Two out of the six objects with such spectra are well-studied
radio sources\footnote{We give the angular sizes of the radio
sources according to the FIRST \cite{backer:Soboleva_n} catalog and their
r-band magnitudes, according to the SDSS survey \cite{adelman:Soboleva_n}. We
give the major axis of the ellipse inferred from the Gaussian
model of the radio source.}.

Dallacasa et al.~\cite{dal1:Soboleva_n} identify the GPS radio
source 152114+043020 (or 4C+04.51) with a z $=$ 1.3 galaxy with a
compact binary structure on a parsec-size scale. Stangellini et
al.~\cite{stan:Soboleva_n} classify this source as a quasar. The
maximum in its spectrum is located at 1\,GHz and its magnitude is
$m_{r}=21.2^{m}$.

The 161637+045932 object from the list of bright HFP sources
\cite{dal2:Soboleva_n} is a quasar with a redshift \mbox{z $=$
3.22}, angular size 0.76$''$, \mbox{$m_{r}=19.3^{m}$}, and
spectral maximum located at about $5$~GHz. Tinti et
al.~\cite{ti2:Soboleva_n} classify this source as a blazar rather
than a young object.

For the remaining four objects we found no published evidence
indicative of their classification with a certain group. These
include two starlike objects: 114631+045818 (with a radio size of
2.27$''$, a spectral maximum at about $5$\,GHz, and
\mbox{$m_{r}=19.9^{m}$)} and 101940+044359  (with a radio size of
3.36$''$ and \mbox{$m_{r}=20.1^{m}$}) and two galaxies:
125755+045917 (90$''$, \mbox{$m_{r}=19.1^{m}$)} and 091636+044132
(94.2$''$, \linebreak \mbox{$m_{r}=16.8^{m}$}, z $=$ 0.18). In
fig.~\ref{fig11:Soboleva_n} one can find the spectra of the
sources 101940+044359, \linebreak 152114+043020, and
161637+045932.

No low-frequency fluxes are known so far for other objects with
the maxima in the spectra. One of them, 151141+051809, has a
spectral maximum at the frequency about $12$~GHz with a flux of
1~Jy and is a yet another HFP-type source
\cite{dal2:Soboleva_n,ti2:Soboleva_n}. It is a nearby bright
($m_{r}=16.3^{m}$) Sy1-type galaxy with a redshift of  z $=$ 0.08.

The other objects of the fourth group are fainter. Their spectral
maxima vary from 1.4 to 8~GHz. Optically  these sources are
identified with quasars, starlike objects and with galaxies. Only
one of these objects was found in an empty field (EF). For quasars
and bright nearby galaxies their spectroscopic redshifts are
known. One of these objects is a blazar 154418+045822 (RX
J15442+0458) with an angular size of 1.54$''$ and
$r_{m}=18.3^{m}$~\cite{laur:Soboleva_n}.

In the Australian Southern-sky catalog at \linebreak
20~GHz~\cite{mas1:Soboleva_n} the number of objects of the fourth type with
spectral maxima in the \mbox{4--8\,GHz} interval exceeds 20\% of
the total number of sources. The detection level of the catalog is
0.5~Jy. Massaridi et al.~\cite{mas2:Soboleva_n} used the results of
observations of the Australian telescope at 4.8 and 8.6\,GHz and
the data from other catalogs at the frequencies of 1.4 and
0.84~GHz. The fraction of sources of the fourth type in this
catalog \cite{mas2:Soboleva_n} is several (about five) times greater than in
our catalog. Similarly to the RCR catalog case, this discrepancy
cannot be explained by selection.

Four sources of the RCR catalog \linebreak (103846+051229,
112417+045144, 115347+045858 and 133921+050159) have spectra that
can be approximated fairly well by a parabolas in the \linebreak
\mbox{1.4--4.8~Ghz} frequency interval. Figure~\ref{fig12:Soboleva_n} (the
upper panel) presents the spectrum of 103846+051229. It is a
triple (according to FIRST data) starlike object
\mbox{($m_{r}=19.1^m$)} with a faint envelope in its SDSS image.

The 112417+045144 source consists of a nucleus with a jet and is
locate in an empty field. The triple source 115347+045858 is
identified with a nearby galaxy (z $=$ 0.23, $m_{r}=17.2^m$). The
point source 133921+050159 is identified with a quasar (z $=$
1.36, $m_{r}=19.1^m$). In these three sources the nucleus
contributes significantly to the overall flux.

Objects with a minimum in the spectrum are called upturn sources
\cite{dent:Soboleva_n}. Tucci et al.~\cite{tucci:Soboleva_n} report the results of the
spectral properties analysis for the sources of an almost complete
VSA sample (brighter than 10~mJy at 15~GHz) in the 1.4--33\,GHz
frequency interval.

The spectra of most of the sources with steep spectra at
1.4--5\,GHz proved to become flat with increasing frequency, and
the fraction of sources with rising spectra at frequencies above
5~GHz in equal to 19\% of the sample. In our list the former make
up for 10\%, and the latter (upturn)---only for $0.01$\%.

This class of sources will possibly be recorded in the future high-resolution observations of the
CMB (Cosmic Microwave Background) \cite{tucci:Soboleva_n}.

Figure~\ref{fig12:Soboleva_n} (the lower panel) demonstrates an example of
the variable source (var) spectrum. Based on the available data,
we classified seven sources as variable; all of them have flat
spectra. The 073357+045614 source is possibly variable and it has
an inverse spectrum. These are mostly sources dominated by the
flux from the nucleus, which are identified with quasars, and some
of them (083148+042938~\cite{sbaru:Soboleva_n}, 123932+044305~\cite{nandi:Soboleva_n},
and 142409+043451~\cite{chen:Soboleva_n}) are classified as blazars. The
101603+051303 source is in the HPF list \cite{dal2:Soboleva_n}.

We already pointed out above that about half of all RCR objects
have flux data available only at two frequencies 1.4  and 3.94~GHz
(plus flux estimates based on  GB6 and VLSS maps). These are faint
objects with fluxes that are mostly below 30~mJy
(Fig.~\ref{fig5:Soboleva_n}b). About 30\% of these objects have power-law
spectra with spectral indices ranging from --1 to --0.5; about
45\% of all objects have flat spectra, and  25\% objects have
inverted spectra. The latter, possibly, have their spectral maxima
higher 10~GHz.

\section{CONCLUSIONS}

We used RATAN-600 observations of the sky band at the declination
$Dec \sim 5\degr$ in the right-ascen\-sion interval $7^h \le$ R.A.
< $17^h$ in 1987--1999 combined with reprocessing of the ``Cold''
survey data (1980--1981) to compile a list (the RCR catalog) of
550 objects identified with objects of the NVSS catalog. This list
includes 18 blends and 15 double sources. Data reduction was
performed using two independent methods.

We determined the fluxes, right-ascensions, and spectral indices
of every object of our list and constructed the histograms of
spectral indices and fluxes for different samples of sources
\footnote{We did not determine the declinations of the sources but
adopted them from the NVSS catalog.}.

We reconstructed the spectra  using all the catalogs available
from the  CATS, Vizier, and NED databases and the flux estimates
based on the maps of the  VLSS and GB6 surveys. These estimates
are useful primarily for the reconstruction of the sources spectra
for that fluxes are known only at two frequencies: 3.94~GHz (RCR)
and 1.4~GHz (NVSS). Such objects make up for about 50\% of the RCR
catalog (245 objects).

These are mostly sources with fluxes not exceeding $30$\,Jy, about
65\% of them have flat or inverse spectra ($\alpha > -0.5$). The
histograms of the fluxes and spectral indices and the average
spectral index of this sample of sources are affected by
selection, since objects with steep spectra cannot be detected due
to the limited sensitivity of the survey.

We analyzed the reliability of the results obtained. We
demonstrated that use of two different methods of data reduction
yields more accurate results both for the right ascensions and
fluxes of the sources. We therefore believe that the technique of
independent reduction of observational data has proved to be
\mbox{successful.}

Note in conclusion that the study of spectral indices at
centimeter-wave frequencies is closely linked to the problem of
eliminating selection effects. For such studies, i.e., for the
complete analysis of the spectra of objects in deep decimeter-wave
sky surveys (NVSS \cite{co1:Soboleva_n}, FIRST \cite{backer:Soboleva_n}), the
sensitivity at centimeter-wave frequencies should be one to two
orders of  magnitude higher than the sensitivity of decimeter-wave
surveys, i.e., at the level of or better than several tens of
$\mu$Jy. So far, such a high sensitivity at centimeter-wave
frequencies could have been achieved only within small sky areas.

The aim of sky surveys made with RATAN-600 radio telescope
\cite{h:Soboleva_n,bu:Soboleva_n} is to obtain more comprehensive information about the
spectral indices of decimeter-wave sources. These surveys serve as
an intermediate link  between deep  VLA surveys and
low-sensitivity all-sky surveys. The main and very important
conclusion of such surveys is that we found no objects within the
right ascension interval considered---at least at the 10--15~mJy
level---that had not been previously included into decimeter-wave
catalogs.

Ninety per cent of the  RCR sources identified with NVSS objects
and having fluxes  ${\rm F_{1.4}} \ge $ 100 mJy lie in the $\Delta
Dec = \pm6'$ band of the survey. In the same band 72\% and 16\% of
the sources were identified with NVSS objects with fluxes \mbox{20
mJy < ${\rm F_{1.4}}$<100 mJy} and ${\rm F_{1.4}} \le $ 20 mJy,
respectively. In the broader band \mbox{$\Delta Dec = \pm20'$} the
number of sources identified with  NVSS objects decreases down to
70\%, 37\%, and 6\%, respectively.

Thus so far for the small population of  NVSS objects all the
centimeter-wave surveys including ``Cold'' and RZF provide the
data mostly for objects with synchrotron self absorption (BL Lac,
QSR, AGN). The  new epoch in this field is expected to start the
operation of ALMA and SKA instruments at millimiter and about 1-cm
waves, respectively.

The catalog is available at \linebreak {\tt
http://cdsavc.u-strasbg.fr./viz-bin/\linebreak
cat?J/other/AstBu/65.42}.

\begin{acknowledgments}

This work was supported in part by the Russian Foundation for
Basic Research (grant nos.~08-02-00486a and 09-07-00320) and the
``Scientific schools'' program of the Russian Academy of Sciences.
We are grateful to A.~I.~Kopylov for his critical comments
concerning the RC catalog.

\end{acknowledgments}


\begin{thebibliography}{99}

\bibitem{pa1:Soboleva_n}
\refitem{article}
Yu.~N.~Parijskij, N.~N.~Bursov, N.~M.~Lipovka, et al.,
\aas~ {\bf87}, 1 (1991).

\bibitem{h:Soboleva_n}
\refitem{misc} Yu.~N.~Parijskij and D.~V.~Korol'kov, {\it Itogi
Nauki i Tekhniki. Ser. Astronomiya} (VINITI, Moscow, 1986) \textbf
{31},  73 (1986).

\bibitem{ber:Soboleva_n}
\refitem{article} A.~B.~Berlin, E.~V.~Bulaenko, V.~Ya.~Gol'nev, et
al., \pazh~ {\bf7},  290 (1981).

\bibitem{pa2:Soboleva_n}
\refitem{article}
Yu.~N.~Parijskij and D.~V.~Korolkov,
Sov. Sci. Rev. Astrophys. Space Rhys.  {\bf5}, 39 (1986).

\bibitem{co1:Soboleva_n}
\refitem{article}
J.~J.~Condon, W.~D.~Cotton, E.~W.~Greisen, et al.,
\aj~  {\bf115}, 1693 (1998).

\bibitem{fr:Soboleva_n}
\refitem{article}
R.~L.~White, R.~H.~Becker, D.~J.~Helfand, and M.~D.~Gregg,
\apj~ {\bf 475}, 479 (1997).

\bibitem{zh:Soboleva_n}
\refitem{misc} O.~P.~Zhelenkova, Candidate's Dissertation in
Mathematics and Physics (Special Astrophysical Observatory of the
Russian Academy of Sciences, Nizhnij Arkhyz, 2007).

\bibitem{so1:Soboleva_n}
\refitem{article} N.~S.~Soboleva, N.~N.~Bursov, and
A.~V.~Temirova, \azh~ {\bf83},  387 (2006).


\bibitem{e1:Soboleva_n}
\refitem{article}
N.~A.~Esepkina, N.~L.~Kaidanivskii, B.~V.~Kuznetsov, et al.,
Radiotekhnika i Elektronika {\bf 6}, 1947 (1961).

\bibitem{e2:Soboleva_n}
\refitem{article} N.~A.~Esepkina, N.~S.~Bakhvalov, B.~A.~Vasil'ev,
et al., Astrofiz.\ Issledovaniya\ (Izv.\ Spec. Astrofiz. Obs.)
{\bf 11}, 182 (1979).

\bibitem{e3:Soboleva_n}
\refitem{article} N.~A.~Esepkina, B.~A.~Vasil'ev, I.~A.~Vodovatov,
and M.~G.~Vysotskij, Astrofiz.\ Issledovaniya\ (Izv.\ Spec.
Astrofiz. Obs.) {\bf 11}, 197 (1979).

\bibitem{m1:Soboleva_n}
\refitem{article}
E.~K.~Majorova,
\bsao~ {\bf 53}, 78 (2002).

\bibitem{m2:Soboleva_n}
\refitem{article}
E.~K.~Majorova and S.~A.~Trushkin,
\bsao~ {\bf 54}, 89 (2002).

\bibitem{m3:Soboleva_n}
\refitem{article}
E.~K.~Majorova and N.~N.~Bursov,
\ab~ {\bf 62}, 398 (2007).

\bibitem{na:Soboleva_n}
\refitem{url}
Skyview,
{\tt http://skyview.gsfc.nasa.gov/cgi- \linebreak -bin/skvbasic.pl}.

\bibitem{m4:Soboleva_n}
\refitem{article}
E.~K.~Majorova,
\ab~ {\bf 63}, 56 (2008).

\bibitem{bu:Soboleva_n}
\refitem{article} N.~N.~Bursov, Yu.~N.~Pariiskii, E.~K.~Maiorova,
et al., \azh~ {\bf84}, 1 (2007).

\bibitem{so2:Soboleva_n}
\refitem{misc} N.~S.~Soboleva, A.~V.~Temirova, and N.~N.~Bursov,
Report of the St. Petersburg Branch of the Special Astrophysical
Observatory of the Russian Academy of Sciences, No. 2 (2008).

\bibitem{mi:Soboleva_n}
\refitem{url}
G.~Miley and ~C.~De Breuck, astro-ph/0802.2770

\bibitem{v1:Soboleva_n}
\refitem{misc} O.~V.~Verkhodanov, S.~A.~Trushkin, H.~Andernach,
and V.~N.~Chernenkov, in {\it Astronomical Data Analysis Software
and Systems VI}, ed. by G. Hunt and H. E. Payne. ASP Conference
Series {\bf 125}, 322 (1997) ({\tt http://cats.sao.ru}).

\bibitem{v2:Soboleva_n}
\refitem{article}
O.~V.~Verkhodanov, S.~A.~Trushkin, H.~Andernach, and V.~N.~Chernenkov,
\bsao~ {\bf 58}, 118 (2005).

\bibitem{viz:Soboleva_n}
\refitem{article}
F.~Ochsenbein, ~P.~Bauer, and ~J.~Marcout,
\aas~ {\bf143}, 23 (2000).

\bibitem{ned:Soboleva_n}
\refitem{url}
NED,
{\tt http://nedwww.ipac.caltech.edu/}.

\bibitem{cohen:Soboleva_n}
\refitem{article}
A.~S.~Cohen, ~W.~M.~Lane, ~W.~D.~Cotton, et al.,
\aj~ {\bf134}, 1245 (2007).

\bibitem{gregory:Soboleva_n}
\refitem{article}
P.~C.~Gregory, ~W.~K.~Scott, ~K.~Douglas and ~J.~J.~Condon,
\apjs~ {\bf103}, 427 (1996).

\bibitem{ode:Soboleva_n}
\refitem{article} C.~P.~O'Dea, \pasp
%Publications of the
%Astronomical Society of the Pacific (PASP)
{\bf 110}, 493 (1998).

\bibitem{ti1:Soboleva_n}
\refitem{article}
S.~Tinti and G.~De Zotti,
\aaa~ {\bf445}, 889 (2006).

\bibitem{ti2:Soboleva_n}
\refitem{article}
S.~Tinti, D.~Dallacasa, G.~De Zotti, et al,
\aaa~ {\bf432}, 31 (2005).

\bibitem{backer:Soboleva_n}
\refitem{article}
R.~H.~Becker, D.~J. Helfand, and R.~L.~White,
\aj~ {\bf475}, 479 (1997).

\bibitem{adelman:Soboleva_n}
\refitem{article}
J.~K.~Adelman-McCarthy  et al., \apjs~ {\bf175}, 297 (2008).

\bibitem{dal1:Soboleva_n}
\refitem{article}
D. Dallacasa,~M. Bondi,~W.~Alef, and F.~Mantovani,
\aas~ {\bf129}, 219   (1998).

\bibitem{stan:Soboleva_n}
\refitem{article}
C.~P.~Stanghellini et al.,
\aaa~ {\bf443}, 891 (2005).

\bibitem{dal2:Soboleva_n}
\refitem{article}
D.~Dallacasa, C.~Stanghellini, M.~Centoza, and R.~Fanti, \aaa~ {\bf363}, 887 (2000).

\bibitem{laur:Soboleva_n}
\refitem{article}
S.~A.~Laurent-Muehleisen et al.,
\aj~ {\bf525}, 127 (1999).

\bibitem{mas1:Soboleva_n}
\refitem{url}
M.~Massaridi, R.~D.~Ekers, T.~Murphy, et al.,
astro-ph/0709.3485.
%arXiv: 0709.3485v3 [astro-ph] 12 Feb 2008

\bibitem{mas2:Soboleva_n}
\refitem{article}
M.~Massaridi, R.~D.~Ekers, T.~Murphy  et al.,
\mnras~ (in press).

%"The Australia Telescope 20GHz (AT20G) Survey: The Bright Source Sample"
%M.Massaridi R.D.Ekers,T.Murphy, R.Ricci, E.M.Sadler,G.De Zotti et al.
%будет напечатана в Mon.Not.R.Astron. в апреле 2009

\bibitem{dent:Soboleva_n}
\refitem{article}
W.~A.~Dent and ~F.~T.~Haddock, Nature {\bf205}, 487 (1965).

\bibitem{tucci:Soboleva_n}
\refitem{article}
M.~Tucci, J.~A.~Rubio-Martin, R.~Rebolo et al.,
\mnras~ {\bf386}, 1729 (2008).

\bibitem{sbaru:Soboleva_n}
\refitem{article}
B.~Sbarufatti, A.~Treves, and R.~Falomo, \aj~ {\bf 635}, 173 (2005).

\bibitem{nandi:Soboleva_n}
\refitem{article}
G.~Nandikotkur et al.,
\aj~ {\bf 657}, 706 (2007).

\bibitem{chen:Soboleva_n}
\refitem{article}
P.~S.~Chen, H.~W.~Fu, and Y.~F.~Gao, New Astronomy {\bf 11}, 27 (2005).

\end{thebibliography}
\end{document}